\documentclass[aps,prb,twocolumn,superscriptaddress]{revtex4-2}

\usepackage{braket}
\usepackage{amsmath}
\usepackage{graphicx}
\usepackage{dcolumn}
\usepackage{bm}
\usepackage{float}
\graphicspath{{Figures/}}

\usepackage{color}
\definecolor{blue}{rgb}{0.1, 0.1, 1}

\def\be{\begin{equation}}
\def\ee{\end{equation}}
\def\bea{\begin{eqnarray}}
\def\eea{\end{eqnarray}}
\usepackage[colorlinks,bookmarks=false,citecolor=blue,linkcolor=black,urlcolor=blue]{hyperref}
\begin{document}

\title{Dimensional crossover and phase transitions in coupled chains: Density matrix renormalization group results}

\author{Gunnar Bollmark}
\affiliation{
	Department of Physics and Astronomy, Uppsala University, Box 516, S-751 20, Uppsala, Sweden
}
\author{Nicolas Laflorencie}
\affiliation{Laboratoire de Physique Th\'eorique, CNRS and Universit\'e de Toulouse, 31062 Toulouse, France}
\author{Adrian Kantian}
\affiliation{
	Department of Physics and Astronomy, Uppsala University, Box 516, S-751 20, Uppsala, Sweden
}

\begin{abstract}
Quasi-one-dimensional (Q1D) systems, i.e., three- and two-dimensional (3D/2D) arrays composed of weakly coupled one-dimensional lattices of interacting quantum particles, exhibit rich and fascinating physics. They are studied across various areas of condensed matter and ultracold atomic lattice-gas physics, and are often marked by dimensional crossover as the coupling between one-dimensional systems is increased or temperature decreased, i.e., the Q1D system goes from appearing largely 1D to largely 3D. Phase transitions occurring along the crossover can strongly enhance this effect. Understanding these crossovers and associated phase transitions can be challenging due to the very different elementary excitations of 1D systems compared to higher-dimensional ones. In the present work, we combine numerical matrix product state (MPS) methods with mean-field (MF) theory to study paradigmatic cases of dimensional crossovers and the associated phase transitions in systems of both hard-core and soft-core lattice bosons, with relevance to both condensed matter physics and ultracold atomic gases. We show that the superfluid-to-insulator transition is a first order one, as opposed to the isotropic cases and calculate transition temperatures for the superfluid states, finding excellent agreement with analytical theory. At the same time, our MPS+MF approach keeps functioning well where the current analytical framework cannot be applied. We further confirm the qualitative and quantitative reliability of our approach by comparison to exact quantum Monte Carlo calculations for the full 3D arrays.
\end{abstract}

\maketitle

\section{\label{sec:level1}Introduction}
Quasi-one-dimensional (Q1D) systems, 3D arrays of weakly coupled 1D quantum systems appear in a wide variety of solid state materials and can readily be realized in lattice-confined ultracold atomic gases. On the materials side, there is very active research into weakly coupled spin chains and ladders such as BPCB~\cite{Klanjsek2008,Bouillot2011} and related magnetic  compounds~\cite{ruegg_thermodynamics_2008,jeong_attractive_2013,blinder_nuclear_2017}, the organic Bechgaard and Fabre salts (``the organics'')~\cite{schwartz_on-chain_1998,Giamarchi2004,Bourbonnais2007,Jerome2016,Kantian2019a}, the strontium-based telephone number compounds~\cite{Nagata1998,Dagotto1999} and chromium pnictide~\cite{Bao2015,Watson2017}, all three of which are itinerant systems which can be made to enter an unconventionally superconducting (USC) state. Of these, the organics, preceding the high-$T_c$ cuprate superconductors as the first USC materials~\cite{Jerome1980}, have received the most in-depth research. Much of this is due to the abiding challenge of resolving the microscopic origin of repulsion-mediated electron pairing as well as the direct transition between the USC state based on this pairing and an insulating magnetically ordered one, analogous to that found in the cuprates, which is of first-order type in the organics~\cite{Jerome2016}. The fascination of the organics is further enhanced by their exhibition of {\it dimensional crossover} (DC), shown by various quantum spin systems as well~\cite{ruegg_thermodynamics_2008,giamarchi_coupled_1999,jeong_magnetic-order_2017,dupont_dynamical_2018}, where the systems effective dimensionality increases from 1D to 2D and eventually 3D, as quantum coherence between the constituent 1D systems increases with decreasing temperature and/or increased (but still weak) intersystem coupling. These DCs can further be marked by a phase transition occurring along the crossover, where DC can then be particularly sharp; for example, the opening of a gap in each constituent 1D system can make it much harder for intersystem coupling to establish coherence and thus ordering in the transverse direction will be much weaker.

The concept of DC taking place around a phase transition is especially interesting for the theory of the USC state, as it is in Q1D models alone that the transition into a superconducting state based on repulsively mediated pairing of fermions can be understood at the fundamental level, at least qualitatively. The prime models for this are 3D arrays of doped, weakly coupled Hubbard ladders. Here, fusing Tomonaga-Luttinger-liquid (TLL) theory for the single ladder~\cite{BookGiamarchi2003} with either static mean-field (MF) theory~\cite{Karakonstantakis2011} or alternatively renormalization group treatments~\cite{Giamarchi2004} allows a qualitative description of the transition to the USC state as the system crosses over from effectively uncoupled 1D Hubbard ladders to the 3D ordered array as temperature decreases.

Going from such a fundamental, qualitative description of the phase transition marking this specific DC to one allowing for quantitative accuracy has stayed an open challenge for which the theoretical tools remain to be developed. Understanding DC and their associated transitions in general with quantitative and even qualitative theory can be very difficult, such as in the case of the organics, because the basis of collective density excitations used to describe the 1D system are completely different from the one of Landau or Bogoliubov quasiparticles used for 2D and 3D systems.

The present work is thus motivated by the twin challenge of developing better theory for DC and associated phase transitions in Q1D-systems in general, as well as specifically for the case of the transition into the USC phase in the Hubbard ladder array. As a first step towards this end we set up a comparatively simple model, 3D arrays formed from weakly coupled chains of interacting lattice bosons with short-range interactions (c.f. Fig.~\ref{hcb_chain_unfilled}). As will be shown and discussed, this class of models combines several advantages: {\it (i)}~They show multiple interesting DCs and associated phase transitions, including first-order transitions (like the organics do between the USC and a magnetically ordered insulating phase) and possibly mixed-order transitions. {\it (ii)}~They are perfect testbeds to further advance efficient yet remarkably accurate numerics based on combinations of matrix product states (MPS) and MF pioneered, e.g., in Ref.~\citenum{Klanjsek2008}. Crucially, the accuracy of these MPS+MF numerics can be ascertained by the gold standard for 2D/3D lattice bosons, quantum Monte Carlo (QMC) simulations. {\it (iii)}~Our MPS+MF numerics can be checked directly against fit-free TLL+MF analytical theory used for the thermal transition to the superfluid regime~\cite{Ho2004,Cazalilla2006a}. Additionally, our numerics will work in regimes where TLL+MF is no longer applicable as well as making possible efficient real-time many-body dynamics for Q1D systems. {\it (iv)}~When specializing the study model to the case of hard-core bosons (HCBs) with nearest-neighbor (n.n.) repulsion, it admits mapping to Q1D arrays of doped Hubbard ladders at the level of low-energy, long-wavelength effective TLL theory. {\it (v)}~These systems either already admit realization in many existing experiments on ultracold lattice gases, including the possibility of observing mixed-order DC, or, in the case of HCBs with n.n. repulsion, may do so within the foreseeable future~\cite{Frisch2015,Baier2018}.

The present paper is thus structured as follows: Section~\ref{sec:model} describes the Q1D array model of bosonic chains and introduces the transverse MF approximation. Section~\ref{sec:methods} describes the MPS+MF method we use for fast, efficient calculations of the systems properties for ground and thermal states. Details of the QMC calculations are also given. Section~\ref{sec:results} discusses the zero-temperature first-order transition we find between a 3D superfluid (SF) and a 1D charge-ordered (CO) phase for HCBs with increasing n.n. repulsion, and the similar transition observed for softcore bosons at integer filling. We also study the transition between SF and a thermal gas with rising temperature. The results of the MPS+MF approach are compared against both QMC and TLL+MF analytics and found to range from excellent to highly satisfactory. In Sec.~\ref{sec:discussion} we summarize the validity of the MPS+MF approach to phase transitions in bosonic systems and discuss the implications of our results for DC physics in other systems as well as consider the efficiency of MPS+MF in comparison with QMC. Section~\ref{sec:conclusion} then provides an outlook on future research on the basis of the present work.

\section{Model}\label{sec:model}
In this work, we consider extended Bose-Hubbard models with anisotropic tunneling strength. We first focus on hard-core bosons (HCB), for which the number of allowed particles is restricted to one boson per site. Further, to connect with established experiments we can also lift this restriction of one boson per site and consider the more general case of soft-core bosons (SCB).
\subsection{3-dimensional Hamiltonian}
The full Hamiltonian is given by the expression
\begin{multline} \label{ham}
H_{B} = -t\sum_{\{ \hat{\textbf{R}}_i \}} b^\dagger_{\hat{\textbf{R}}_i+\hat{\textbf{x}}}b_{\hat{\textbf{R}}_i}^{\vphantom\dagger} + \text{h.c.} - \mu\sum_{\{ \hat{\textbf{R}}_i \}} b^\dagger_{\hat{\textbf{R}}_i}b_{\hat{\textbf{R}}_i}^{\vphantom\dagger} \\ + \frac{U}{2}\sum_{\{ \hat{\textbf{R}}_i \}}n_{\hat{\textbf{R}}_i}\left(n_{\hat{\textbf{R}}_i} - 1\right) + V\sum_{\{ \hat{\textbf{R}}_i \}}n_{\hat{\textbf{R}}_i+\hat{\textbf{x}}}n_{\hat{\textbf{R}}_i} \\ - t_\perp\sum_{\{ \hat{\textbf{R}}_i \},~ \hat{\textbf{a}} \in [\hat{\textbf{y}},\hat{\textbf{z}}]} b^\dagger_{\hat{\textbf{R}}_i+\hat{\textbf{a}}}b_{\hat{\textbf{R}}_i}^{\vphantom\dagger} + \text{h.c.} \\ = H_t + H_\mu + H_U + H_V + H_{t_\perp},
\end{multline}
where $\{\hat{\textbf{R}}_i\}$ denotes the set of all lattice points, $b^\dagger\vphantom{n}_{\hat{\textbf{R}}_i}$ ($b_{\hat{\textbf{R}}_i}$) is the creation (annihilation) operator associated with the site at $\hat{\textbf{R}}_i$ and $n_{\hat{\textbf{R}}_i}=b^\dagger\vphantom{n}_{\hat{\textbf{R}}_i}b_{\hat{\textbf{R}}_i}$ is the number operator on that site. We have set the lattice spacing~$a=1$.

The transverse hopping $t_\perp$ governs two directions and the longitudinal hopping one direction. In this paper we consider cases where $t_\perp/t\ll 1$. Further, we restrict ourselves to $U,V>0$, i.e., repulsive interactions. In addition, note that the repulsive interaction $V$ between nearest neighbors only occurs along the strong tunneling direction.

\subsection{Local Hilbert space truncation}
The Hamiltonian Eq.~\eqref{ham} allows any number of bosons on one site: $\braket{n_i}\in[0,\infty]$. In the hard-core case,  $U\to\infty$, such that $0\leq\braket{n_i}\leq 1$. In the soft-core case we let $0\leq\braket{n_i}\leq3$. This cutoff of three bosons per site is chosen such that projections onto states of larger occupation number carry a small weight:
\begin{equation} \label{projection}
	\bra{\Psi}P_4(i)\ket{\Psi} \leq 10^{-4},
\end{equation}
where $P_4(i)$ is the projector onto the state of four bosons on site $i$. For SCB the value of $U$ is of course very important and will be specified, while for simplicity we fix $V=0$ in this model.

\subsection{Quasi-1D Hamiltonian}
We wish to use the density matrix renormalization group (DMRG) algorithm in matrix product state (MPS) formalism \cite{White1992a,Schollwock2011} to solve our problem. However, calculations on 3D models using DMRG scale very poorly with system size. To bypass this issue we reduce the problem to solving an effectively one-dimensional (1D) model using mean-field theory. We consider fluctuations around an order parameter
\begin{equation}
b_{\hat{\textbf{R}}_i} = \left<b_{\hat{\textbf{R}}_i}\right> + \delta b_{\hat{\textbf{R}}_i}
\end{equation}
and ignore terms in the Hamiltonian of order $\mathcal{O}(\delta b^2)$. We make this substitution only in the transverse hopping Hamiltonian $H_{t_\perp}$. If we consider open boundary conditions (OBC) this yields the Q1D Hamiltonian
\begin{multline} \label{smf_ham}
H_{SMF}(\alpha) = -t\sum_{i+1}^{L-1} b^\dagger_{i+1}b_{i}^{\vphantom\dagger}  + \text{h.c.} - \mu\sum_{i=1}^L b^\dagger_{i}b_{i}^{\vphantom\dagger}  \\ + \frac{U}{2}\sum_{i=1}^Ln_{i}\left(n_{i} - 1\right) +  V\sum_{i=1}^{L-1}n_{i+1}n_{i} - \alpha\sum_{i=1}^L \left(b^\dagger_{i} + b_{i}^{\vphantom\dagger} \right),
\end{multline}
where indices $i$ have been introduced which indexes the site of a one-dimensional subset of the 3D model in the longitudinal direction. In this work we will use both OBC and periodic boundary conditions (PBC), the latter in which we have the additional condition of $b_{L+1}=b_1$ and the term 
\begin{equation}
	H_L = -t\left(b^\dagger_1b_{L}^{\vphantom\dagger}  + \text{h.c.}\right)  + Vn_1n_L
\end{equation}
must be added to the Hamiltonian~Eq.~\eqref{smf_ham}.

\begin{figure}[t]
	\flushleft
	(a)
	\includegraphics[width=\columnwidth]{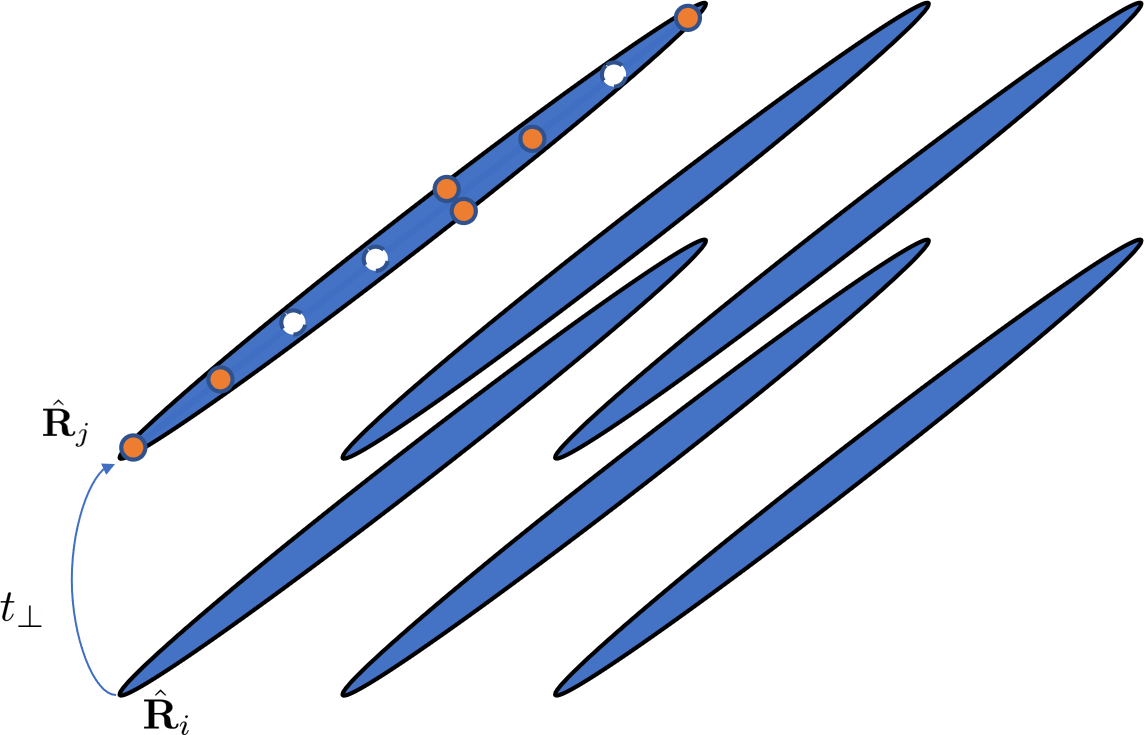}
	
	\vspace{1.0cm}
	(b)
	\includegraphics[width=\columnwidth]{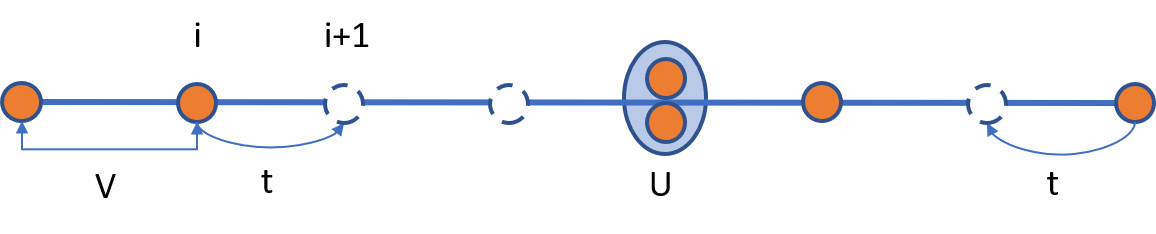}
	\caption{A schematic representation of the model described by Eq.~\eqref{ham} where (a) represents the 3D model and (b) the 1D sub-sets of the full 3D model.}
	\label{hcb_chain_unfilled}
\end{figure}

We only decouple the 3D system transversely since the coupling $t_\perp/t\ll1$ is small by choice. We have routinely ignored any constant contribution to the Hamiltonian. The new coupling $\alpha$ is obtained as
\begin{equation} \label{alp_eq}
\alpha^{(*)} = z_ct_\perp\left<b^{(\dagger)}\right>,
\end{equation}
where $z_c=4$ is the coordination number for a simple cubic lattice and we have assumed that $\alpha$ is real. We will call the constant $\alpha$ a boson injection/ejection amplitude. Notably, the only difference between a 2D and 3D anisotropic system in this approach is $z_c$. A schematic representation of this model is shown in Fig. \ref{hcb_chain_unfilled}.

\section{Methods}\label{sec:methods}
To find ground states and thermal states of the Hamiltonian~Eq.~\eqref{ham} two methods will be used. The first one is comprised of using DMRG to solve the Q1D Hamiltonian~Eq.~\eqref{smf_ham}. We then use quantum Monte Carlo (QMC) simulations~\cite{syljuasen_quantum_2002} for the simplest case of hard-core bosons that we can directly compare with DMRG results.
\subsection{DMRG with static mean-field}
Since~Eq.~\eqref{smf_ham} is a one-dimensional Hamiltonian, the DMRG algorithm scales well with system size and can be used to compute ground states and thermal states~\cite{Schollwock2011}. The additional cost to this method is the self-consistent determination of $\alpha$. We will call the outlined procedure MPS+MF for the remainder of this paper.
\subsubsection{Boson injection convergence}
The self-consistent routine starts with guessing a boson injection amplitude $\alpha_0$ and then computing a new value $\alpha_1$
\begin{equation}
	\alpha_1 = z_ct_\perp\left<b\right>_0,
\end{equation}
where $\left<~\right>_0$ denotes an average with Hamiltonian $H_{SMF}(\alpha_0)$ defined by Eq.~\eqref{smf_ham} with $\alpha=\alpha_0$ and $b$ is in principle any $b_i$ given an infinite system though in practice an average over several sites. Extending the relation to an arbitrary number of loops simply yields
\begin{equation}
\alpha_{n+1} = z_ct_\perp\left<b\right>_n.
\end{equation}
Several exit conditions of the self-consistent loop can be used. Different observables converge at various rates (e.g., density typically converges quickly). In the present case, we use the Bose-Einstein condensate (BEC) order parameter as the observable for determining whether the self-consistent calculation has converged with the condition
\begin{equation}\label{orp_converge}
	\left|\frac{\left<b\right>_n - \left<b\right>_{n-1}}{\left<b\right>_{n-1}}\right| < \epsilon_{\left<b\right>}.
\end{equation}
The quantity $\epsilon_{\left<b\right>}=10^{-4}$ is an error tolerance which can be selected to desired convergence error.

The convergence error should be the largest error in the problem. Any other larger error scale allows $\alpha$ to fluctuate within that scale which disallows settling on a value to convergence precision. An example of such a potential error scale is the truncation error inherent to DMRG.

The fast convergence of this algorithm is highly dependent on how good the initial guess is. Therefore, we have found it good practice to implement some guessing heuristic. For the data shown in this paper, we select an initial value of $\alpha$ which places us above the converged value and check the trend of the computed values of $\alpha$. For the models we consider this trend is usually exponential. Restarting the whole algorithm using the extrapolated value from an exponential fit as initial guess typically brings you closer to the correct value. Thereby the number of loops required for convergence is reduced.

When using this approach we ran into slow-downs of the self-consistent loop convergence close to phase transitions. If no extrapolation scheme as described previously is used, convergence at transition points may require intractably many loops.
\subsubsection{Density targeting}
One issue in the mean-field treatment presented in~Eq.~\eqref{smf_ham} is that the Hamiltonian is transformed from representing a particle number-conserving system to one of nonconservation. Physically, this means that while the particle number is conserved in the full 3D system each individual chain may exchange particles with other chains thus upsetting the conserved particle number locally.

Often we wish to fix the density of an individual chain to some value $n_0$ and must choose the corresponding chemical potential $\mu$. When converging $\alpha$ in the self-consistent loop, the density for one value of $\alpha$ may have different dependence on $\mu$ than for other $\alpha$ s.t.
\begin{equation}
	n_{\alpha_n}(\mu) \not= n_{\alpha_{n+1}}(\mu),
\end{equation}
where 
\begin{equation}
	n_{\alpha_n} = \frac{1}{L}\sum_{i=1}^L\left<n_i\right>_{\alpha_n}.
\end{equation}
This means that in addition to converging $\alpha$ self-consistently we must do the same for $\mu$ simultaneously. This procedure involves the measurement of density each loop and then the calculation of a new chemical potential that gives you the desired density.

Due to this issue the cost of computation increases as each new $\mu$ requires a new state calculation to verify the density, i.e., one self-consistent loop may require several DMRG computations. Fortunately, the density typically converges faster than $\alpha$ and the performance is not greatly affected by the fixation of $\mu$ in the cases considered in this paper.

The density is compared to a chosen target and must fulfill the following condition
\begin{equation}
	\frac{\left|n_{\alpha_{n}} - n_{target}\right|}{n_{\alpha_n}} < \epsilon_{n},
\end{equation}
where $\epsilon_{n}=10^{-5}$ is the error tolerance used for densities in calculations.

\subsection{DMRG observables}
When using DMRG we will consider two observables to characterize the studied phases. The BEC order is evaluated by measuring the expectation values $\left<b_i\right>$. When these are finite it means that there is a finite probability for particles to tunnel in and out of the quasi-1D system described by~Eq.~\eqref{smf_ham}, i.e., the boson injection/ejection amplitude is nonzero. This quantity is our mean-field order parameter and it is computed by averaging over several sites of the quasi-1D model
\begin{equation}\label{orp}
\left<b\right> = \frac{1}{i_l-i_f+1}\sum_{i=i_f}^{i_l}\left<b_i\right>,
\end{equation}
where $i_f$ ($i_l$) is the first (last) site to be included in the average. This calculation assumes that there is no preferential site in the quasi-1D system from which to tunnel in or out. The choice of $i_f,~i_l$ depends on the boundary conditions. This means that OBC requires an average of the systems central sites to avoid boundary effects while PBC is free from this issue as all sites are equivalent.

Typically, DMRG is more efficient with OBC. However, in the hard-core case, OBC gives the system large boundary effects, as is shown in Appendix~\ref{App_OBC}. Thus, the boundary conditions we will use when resolving the SF-CDW transition are PBC for the hard-core case and OBC for the soft-core case.

To characterize CDW phases we compute the charge gap, i.e., the energy required to add or remove one particle from the system. Since the Hamiltonian Eq.~\eqref{smf_ham} does not conserve particle number we will, in this work, define the charge gap as the width in $\mu$ of density plateaus
\begin{equation}
	\Delta_\rho = \mu_{upper}-\mu_{lower},
\end{equation}
where the chemical potentials at the plateau edges are defined by
\begin{flalign}
	&n\left(\mu\right) = \text{const.},~\mu\in\left[\mu_{lower},\mu_{upper}\right] \\
	&n\left(\mu_{upper}+\delta\right) > n\left(\mu_{upper}\right) \\
	&n\left(\mu_{lower}-\delta\right) < n\left(\mu_{lower}\right),
\end{flalign}
where $\delta>0$ is a small addition (subtraction) of the chemical potential. Further details about the charge gap are given in Appendix~\ref{App_plateau}.

\subsubsection{Truncation error extrapolation of DMRG observables} \label{sec:trunc_err}
In order for results from a DMRG solution to be reliable an extrapolation to zero truncation error is required \cite{Schollwock2011}. This is done for all observables $X$ using a linear fit to the data points \cite{White2005}:

\begin{equation}
	X = X_0 + c_0\epsilon_t.
\end{equation}
We find for OBC that this expression fits not only energies but also measurements of the order parameter Eq.~\eqref{orp}.

Using OBC the truncation error is small even for a modest bond dimension as low as $\chi=50$. Extrapolations to zero truncation error yield no improvements within the self-consistent error. On such occasion we do not perform extrapolations and use the largest bond dimension (smallest truncation error) available.

We note that when using PBC quite large bond dimensions are required. When computing charge gaps we have found truncation errors as large as $\epsilon_t\sim 10^{-5}$ for a bond dimension of $\chi=250$. Further, the manner in which charge gaps are computed in this paper carries an additional error (see Appendix~\ref{App_plateau}). This has made extrapolations in truncation error difficult. As a result, the charge gap data in Fig.~\ref{SF_CDW_trans}(a) comes with the caveat that it is affected by notable truncation errors.

\subsubsection{Finite size extrapolation of DMRG observables}
The type of DMRG used in the MPS+MF method is finite size DMRG to make onsite measurements and correlator measurements possible. We are often interested in the thermodynamic behavior of a system and thus we must extrapolate results to the limit of infinitely large systems. The extrapolation scheme used depends on the observable that is being measured.

For the charge gap we use a second degree polynomial fit in $L^{-1}$:
\begin{equation}\label{cgap_fit}
	\Delta_\rho(L) = c_0 + c_1\frac{1}{L} + c_2\frac{1}{L^2} + \mathcal{O}\left(\frac{1}{L^3}\right).
\end{equation}
This expression is commonly used to fit the finite size dependence of energies. We find that our charge gap measurements fit this ansatz as well.

For the order parameter we use two different fitting forms. When used to characterize the finite-temperature second-order normal to superfluid phase transition we use a power-law expression
\begin{equation}\label{orp_fit_1}
	\left<b\right>(L) = c_0 + c_1L^{-c_2}.
\end{equation}
This expression is known to hold analytically at the transition point, and we find our data for finite temperature fits Eq.~\eqref{orp_fit_1} quite well.

For the first order zero-temperature transitions from superfluid to CDW we use a second order polynomial for the squared order parameter
\begin{equation}\label{orp_fit_2}
	\left<b\right>^2 = c_0 + c_1\frac{1}{L} + c_2\frac{1}{L^2} + \mathcal{O}\left(\frac{1}{L^3}\right).
\end{equation}
These expressions hold close to phase transitions which is also the area where finite size effects are most prominent. Concrete examples of such extrapolations are provided in Appendix~\ref{App:fss}.

Frequently, the largest error of the MPS+MF approach is from the self-consistent convergence as opposed to finite size errors. When this occurs, fitting to one of the forms Eqs.~\eqref{cgap_fit}-\eqref{orp_fit_2} is difficult and yields poor fits. On these occasions we find that larger sizes do not change measured value outside of the self-consistent error and we use the largest size measurement available.

\subsection{Quantum Monte Carlo}
Our large-scale QMC simulations have been performed with the stochastic series 
expansion (SSE) algorithm~\cite{syljuasen_quantum_2002} on 3D arrays of coupled chains, using anisotropic lattices of sizes
$L_x$$\times$$L_y\times$$L_z$, with $L_x = L$ and $L_y=L_z=L/f$ for an integer $f$. We have only focused on the case of HCB, but extending to SCF is straightforward. Note also that PBC are used in all directions.

In order to address the bosonic phases and associated transitions for the 3D model Eq.~\eqref{ham} at both zero and finite temperatures, we compute the three following observables.

Charge density wave order is evaluated with the staggered correlation function at mid-distance, along the chain directions
\be
C_{\rm stagg.}=\frac{1}{N}\sum_i (-1)^{L/2}\left(\langle n_i n_{i+L/2}\rangle-\langle n_i\rangle\langle n_{i+L/2}\rangle\right),
\label{eq:CDW}
\ee
where the sum is performed over the $N=L^3/8$ sites and where $f=2$ has been used for the aspect ratio.

The BEC order parameter (condensate density) is obtained by summing off-diagonal correlators
\be
\rho_0=\frac{1}{N^2}\sum_{i,j}\langle b_i^{\dagger} b_j^{\vphantom\dagger}\rangle.
\label{eq:BEC}
\ee
where an aspect ratio of $f=4$ has been used.

The superfluid response can be evaluated for longitudinal (intrachain) and transverse (interchain) directions with the superfluid stiffness 
\be
\rho_{S,\parallel(\perp)}=\frac{1}{N}\frac{\partial^2 E_0(\varphi_{\parallel(\perp)})}{\partial \varphi_{\parallel(\perp)}^{2}}\Big|_{\varphi_{\parallel(\perp)}=0}.
\label{eq:SF}
\ee
In the above definition, $E_0$ is the total energy, and $\varphi_{\parallel(\perp)}$ is a small twist angle enforced on all bonds in both longitudinal and transverse directions. Technically, the superfluid stiffness~\cite{fisher_helicity_1973} is efficiently measured via the fluctuations of the winding number~\cite{pollock_path-integral_1987} during the SSE simulation~\cite{sandvik_finite-size_1997}.

\section{Results}\label{sec:results}
For most of the results we perform three different types of calculations: \textit{(i)} an MPS+MF calculation of both ground and thermal states for hard-core bosons which we also compare with \textit{(ii)} a correspondent QMC calculation and finally \textit{(iii)} calculations of both ground and thermal states for soft-core bosons using MPS+MF.

For MPS+MF, a bond dimension of $\chi=50$ has been used with OBC. For PBC we instead use a bond dimension of $\chi=250$. As stated in Sec.~\ref{sec:trunc_err} truncation error extrapolation has proved difficult since other error sources are dominant. Nevertheless, such extrapolations have been used where possible. We have omitted error bars where the error is smaller than the symbol size in all figures. For all data, the error due to truncation error is much smaller than other sources and we omit such analyses.

First, we analyze the system when interaction strength $U,~V$ is varied. We choose to analyze commensurate densities, where we expect quantum phase transitions to occur at zero temperature. In the hard-core case it is known that charge ordering occurs for a half-filled isolated chain via a Berezinskii-Kosterlitz-Thouless (BKT) transition at $V_c=2t$~\cite{laflorencie_finite_2001}.
This particular density $n = \frac{1}{L}\sum_{i=1}^L\left<n_i\right> = 0.5$ is interesting as it forces the system to incur some energy penalty along with an energy gain from hopping due to the repulsive interaction. We expect there to be a charge-ordering transition for the quasi-1D model as well  but with a shifted $V_c$ compared to the 1D case.

For the soft-core case we instead target $n=1$ and fix $V=0$ to simplify the analysis. This unit-filled regime with only local repulsions can generally be expected to yield some type of order-to-order phase transition~\cite{Bloch2008}.

Second, we analyze the same systems but at finite temperature. We are primarily interested in the critical temperature and how it depends on the microscopic parameters of the Hamiltonian. In this context, we are interested in how accurate our approximate (but numerically low-cost) MPS+MF-based calculations of $T_c$ are in comparison to those from quasiexact QMC.

\subsection{Zero temperature results}
For small values of repulsion we expect there to be a BEC superfluid (SF) phase. At large values of repulsion, the system should become insulating and exhibit a charge-ordered phase (CDW). To analyze this transition we fix $t_\perp/t = 0.05$.

\begin{figure}[t]
	\includegraphics[width=\columnwidth]{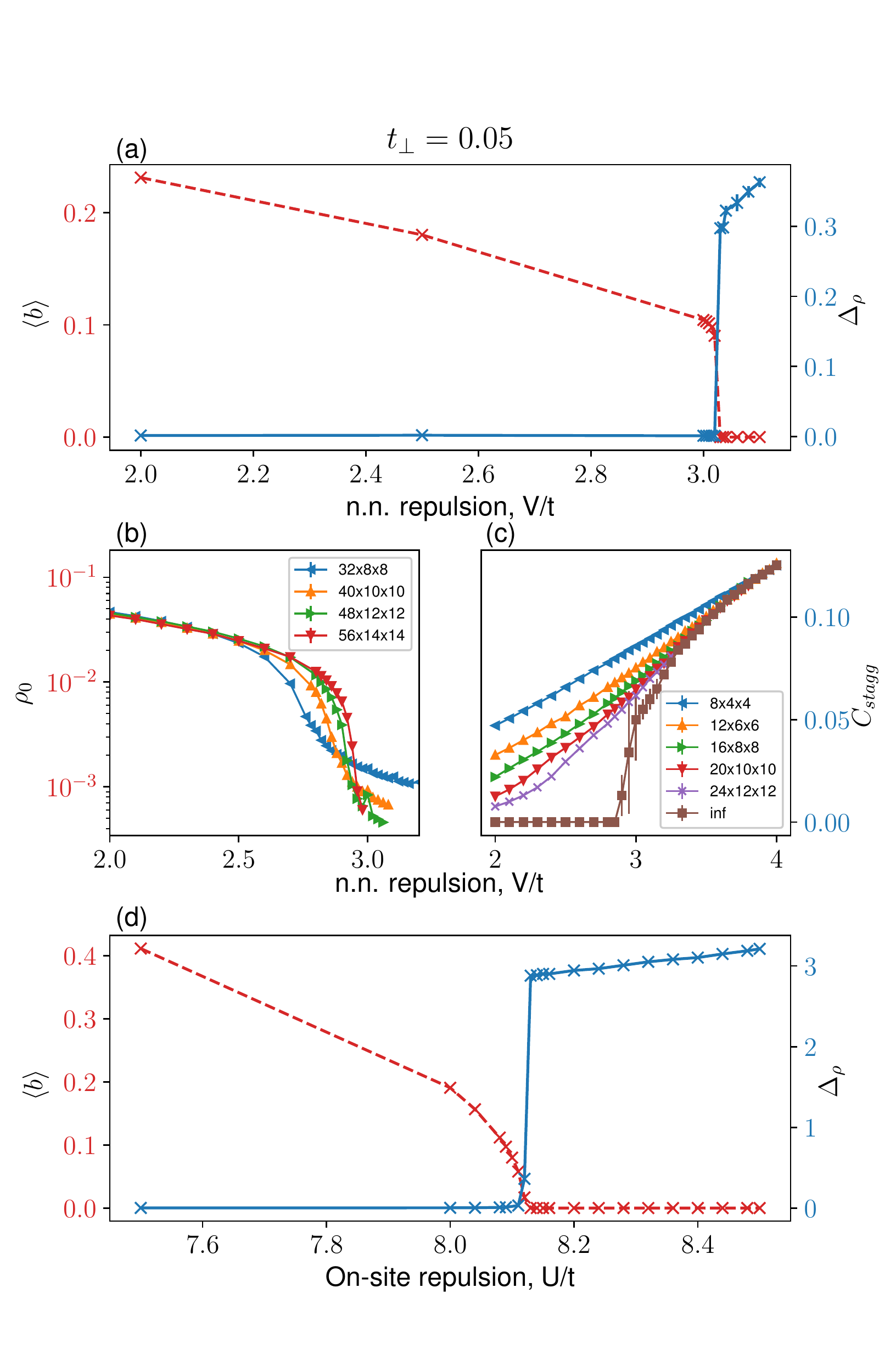}
	\caption{A comparison of the order parameters characterizing the two ordered phases at $T=0$ at $t_\perp/t=0.05$. The red dashed line is the BEC order parameter and the blue solid line the charge gap (which can be seen as the charge density wave order parameter). (a) MPS+MF results for a hard-core constraint model with filling fraction $n=0.5$ using PBC. (b) QMC for hard-core boson model with BEC order parameter Eq.~\eqref{eq:BEC} size dependence. (c) QMC for hard-core boson model with the CDW order parameter Eq.~\eqref{eq:CDW} size dependence. (d) MPS+MF results for a soft-core boson model with filling fraction $n=1.0$ using OBC.}
	\label{SF_CDW_trans}
\end{figure} 

\subsubsection{BEC/Superfluid to CDW at $T=0$}

MPS+MF results are shown in Fig.~\ref{SF_CDW_trans}(a) where the charge gap is plotted together with the BEC order parameter as a function of the nearest-neighbor repulsion~$V$. Note that for an isolated 1D system the transition into CDW occurs at $V/t=2$ whereas in the quasi-1D case we discuss here, the transition is pushed to quite a higher value~$V_c/t\approx3.02$, while $t_\perp/t=1/20$ is small.
Importantly, one observes clear discontinuities for both order parameters at $V_c$, indicating a first-order transition between a gapless BEC-SF and a CDW insulator. The MPS+MF results can be directly compared to the QMC simulations shown in Figs.~\ref{SF_CDW_trans}(b) and \ref{SF_CDW_trans}(c). The agreement is very good, since QMC results find a first-order transition for $V_c/t\approx 3$, the first-order (discontinuous step) character of the transition becoming more and more evident upon increasing system size.

QMC data in Figs.~\ref{SF_CDW_trans}(b) and \ref{SF_CDW_trans}(c) show strong finite size effects, which are more pronounced close to the transition. The BEC density $\rho_0$ [panel (b)] is shown for an aspect ratio of 4. There, $\rho_0(L)$ becomes steeper when increasing system size, a trend which is clearly compatible with a small but finite jump at the thermodynamic limit. This is further discussed in Appendix~\ref{iso_example} where such a jump is more visible due to a larger value of the transverse tunneling. The CDW order parameter $C_{stagg}$, shown in Fig.~\ref{SF_CDW_trans}(c), has been computed for a different aspect ratio of $2$ in order to get a better convergence towards the thermodynamic limit. Using a general finite size scaling of the form
\begin{equation}
C_{stagg}(L)=C_{stagg}^{\infty}+A/L^B\exp(-L/\xi),
\end{equation}
a very good description of finite size data is obtained. The infinite size extrapolation $C_{stagg}^{\infty}$, plotted against $V/t$, is clearly compatible with a jump at the transition. Note however the strong error bars in the critical regime, characteristic of a first order transition.

The soft-core boson data has been computed for $V=0$. Since we fix the density to $n=1.0$ a nearest neighbor repulsion would disturb the potential Mott insulator that can be established at large $U$. In Fig.~\ref{SF_CDW_trans}(c) a transition to the CDW phase can be seen at $U_c/t\approx8.12$. This strongly contrasts with the isolated chain case where a BKT transition occurs for a much smaller onsite repulsion at $U_c/t\approx 3.3$~\cite{krutitsky_ultracold_2016}.

At the transition point the charge gap attains a large value seemingly discontinuously while no such strong first order behavior is apparent when considering the order parameter. It is possible that the latter has a jump so small that it is undetectable by the current method we are using (see Sec.~\ref{sec:discussion}).

\subsection{Finite temperature}
Using the MPS+MF method it is also possible to obtain thermal averages  \cite{Schollwock2011}, while for QMC finite temperature is natural. Thus, we next investigate $T>0$ physics.

\subsubsection{SF to normal}
An interesting transition that should occur for finite temperature is that of 3D superfluid to a thermal gas (the normal or disordered phase). We wish to compute the critical temperature where the system looses BEC coherence and enters the normal phase. We will let the repulsion vary in the system to see how critical temperature is affected. Since we are mainly interested in the SF to normal phase transition we will stay away from values of the repulsion in which there is no SF even at zero temperature, i.e., we stay at $V/t<3$ for hard-core bosons and $U/t<8.12$ for soft-core bosons.

\begin{figure}
	\includegraphics[width=\columnwidth]{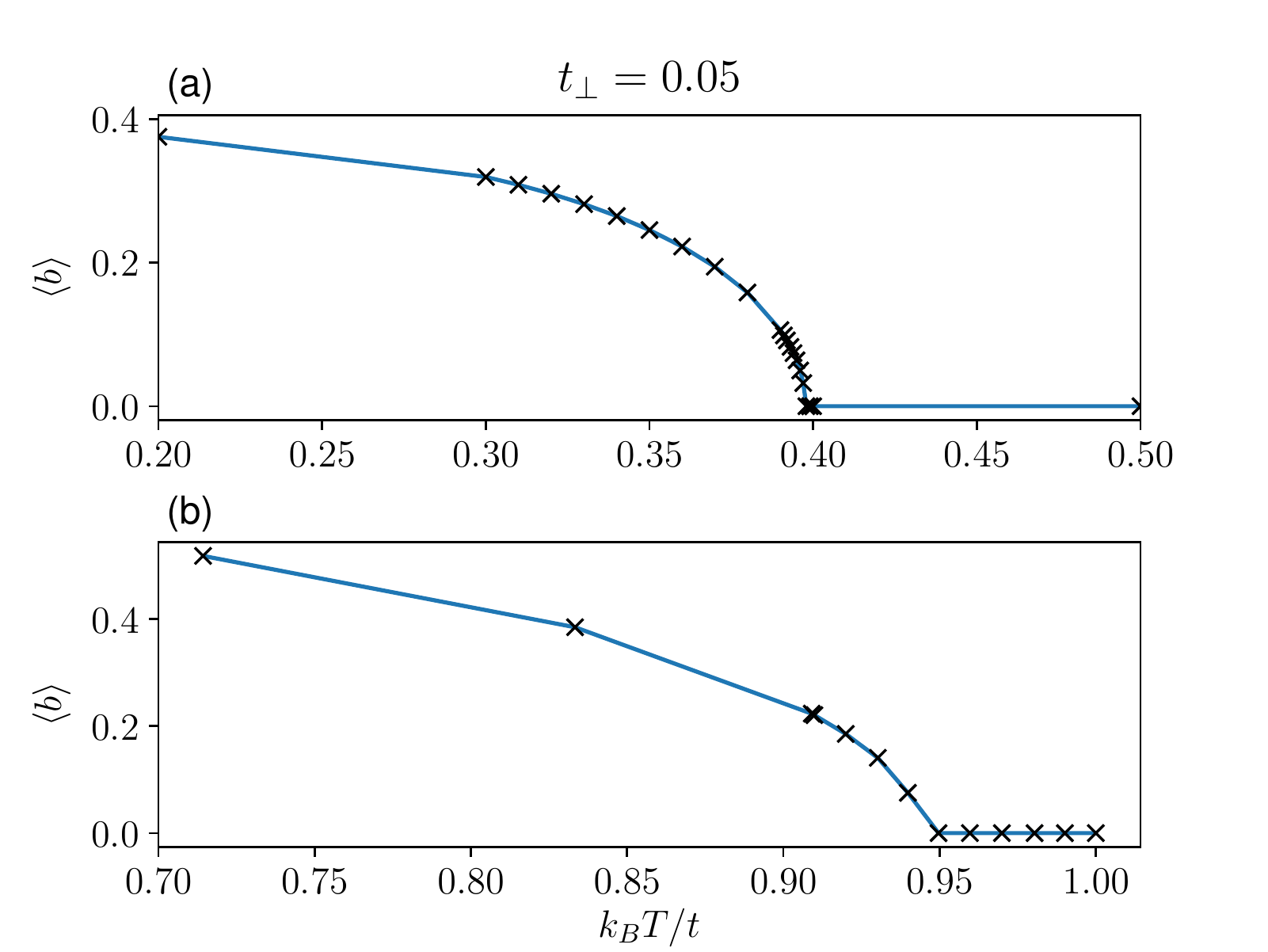}
	\caption{A plot of the superfluid order parameter defined by $\left<b\right> = \frac{1}{i_l-i_f}\sum_{i=i_f}^{i_l}\left<b_i\right>$ extrapolated to infinite longitudinal size versus temperature. (a) Hard-core constraint model with $V=0$ and filling fraction $n=0.5$ using OBC. (b) Soft-core boson model with filling fraction $n=1.0$ and $U/t=6.0$ using OBC.}
	\label{orpvT_hcb}
\end{figure}

The critical temperature of the transition can be found by finding the point at which $\left<b\right>\to0$ in the thermodynamic limit. From Fig.~\ref{orpvT_hcb} it is clear that for $t_\perp/t=0.05$ this point lies close to $T/t=0.4$. An important question using our MPS+MF approach is how accurate the observed critical temperatures are (i.e., how incorrect is the mean-field approximation). A full mean field analysis overestimates the critical temperatures by a factor of $2$ compared to exact calculations using QMC in the 3D case~\cite{Carrasquilla2012}. Hence, it is important to determine if and by how much our MPS+MF hybrid approach improves upon this factor.

\begin{figure}
	\includegraphics[width=\columnwidth]{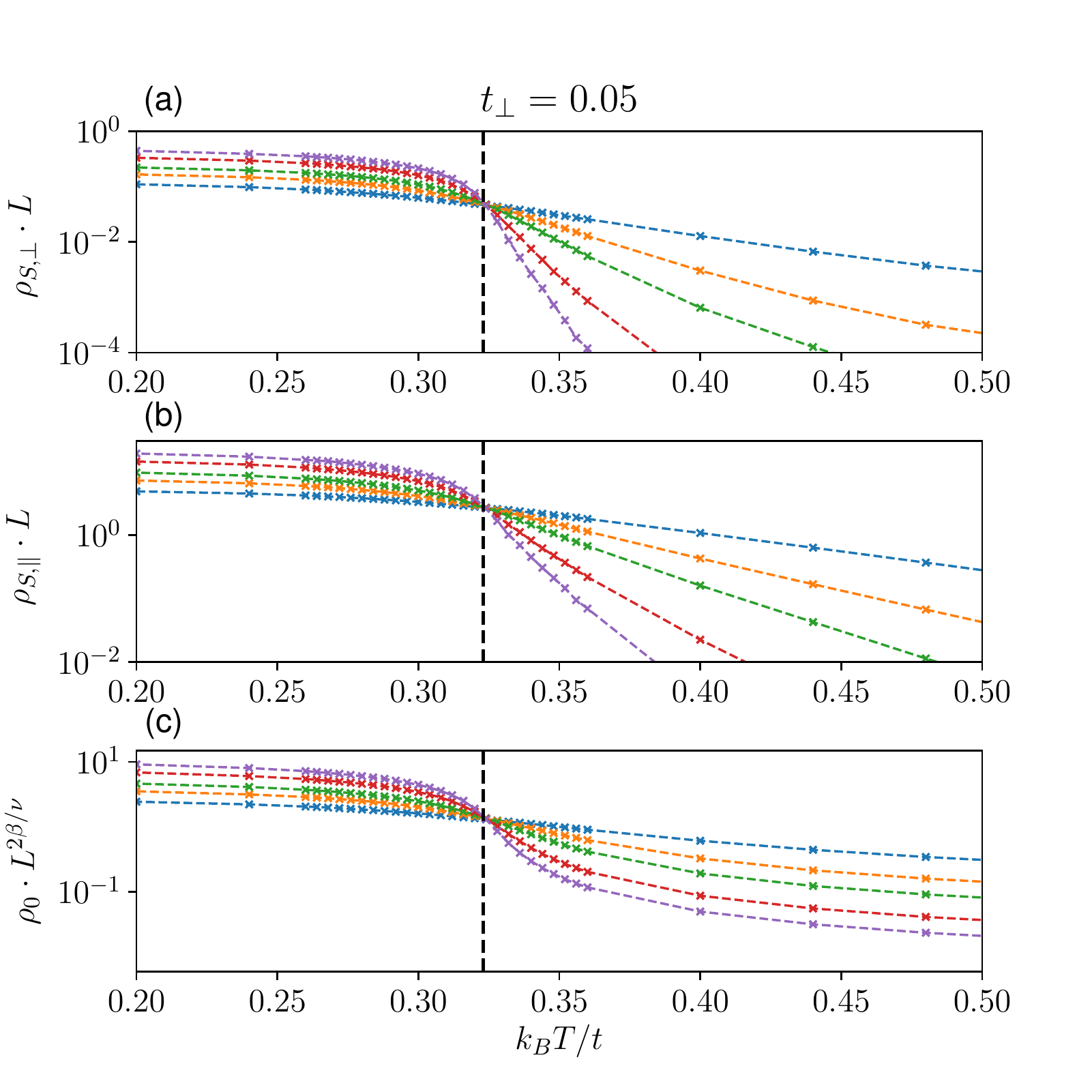}
	\caption{Finite temperature QMC data for the full 3D model Eq.~\eqref{ham} at $V=0$ indicating a transition at $T_c/t=0.323(1)$ using (a) the transverse superfluid stiffness $\rho_{S,\perp}$ scaled with $L$, (b) the longitudinal superfluid stiffness $\rho_{S,\parallel}$ scaled with $L$, (c) the condensate density $\rho_{0}$ scaled with $L^{2\beta/\nu}$ (see text).}
	\label{QMC_orpvT_hcb}
\end{figure}

We have therefore performed finite-$T$ QMC simulations of the full 3D Hamiltonian~Eq.~\eqref{ham}. We determine the critical temperatures using standard finite-size scaling analysis which yields crossings for stiffnesses and BEC order parameter:
\be
\rho_{S,\parallel(\perp)}(T_c)\times L^{z+d-2},
\ee
with $d=3$, $z=0$ for a thermal transition, and
\be
\rho_0(T_c)\times L^{2\beta/\nu},
\ee
where $\beta= 0.3486$ and $\nu=0.6717$ are the critical exponents of the 3D $XY$ universality class~\cite{burovski_high-precision_2006,campostrini_theoretical_2006}. From the results given in Fig.~\ref{QMC_orpvT_hcb} the three crossings are in perfect agreement, giving for $V=0$ a critical temperature $T_c/t=0.323(1)$. Compared to the critical temperature from our MPS+MF approach of $T_c/t\approx0.4$ we find that the difference is significantly better than a factor of $2$\cite{Rigol2005}.

The soft-core model finite temperature data is computed for $U/t=6.0$ since leaving $U$ too small makes the local Hilbert space truncation increasingly erroneous. A notable feature is the increased critical temperature at around $T_c/t\approx0.95$ as seen in Fig.~\ref{orpvT_hcb}(b), which puts these transitions squarely within the range of being observable within current experiments.

\subsubsection{$T_c$ dependence on $t_\perp$}
\begin{figure}[t]
	\includegraphics[width=\columnwidth]{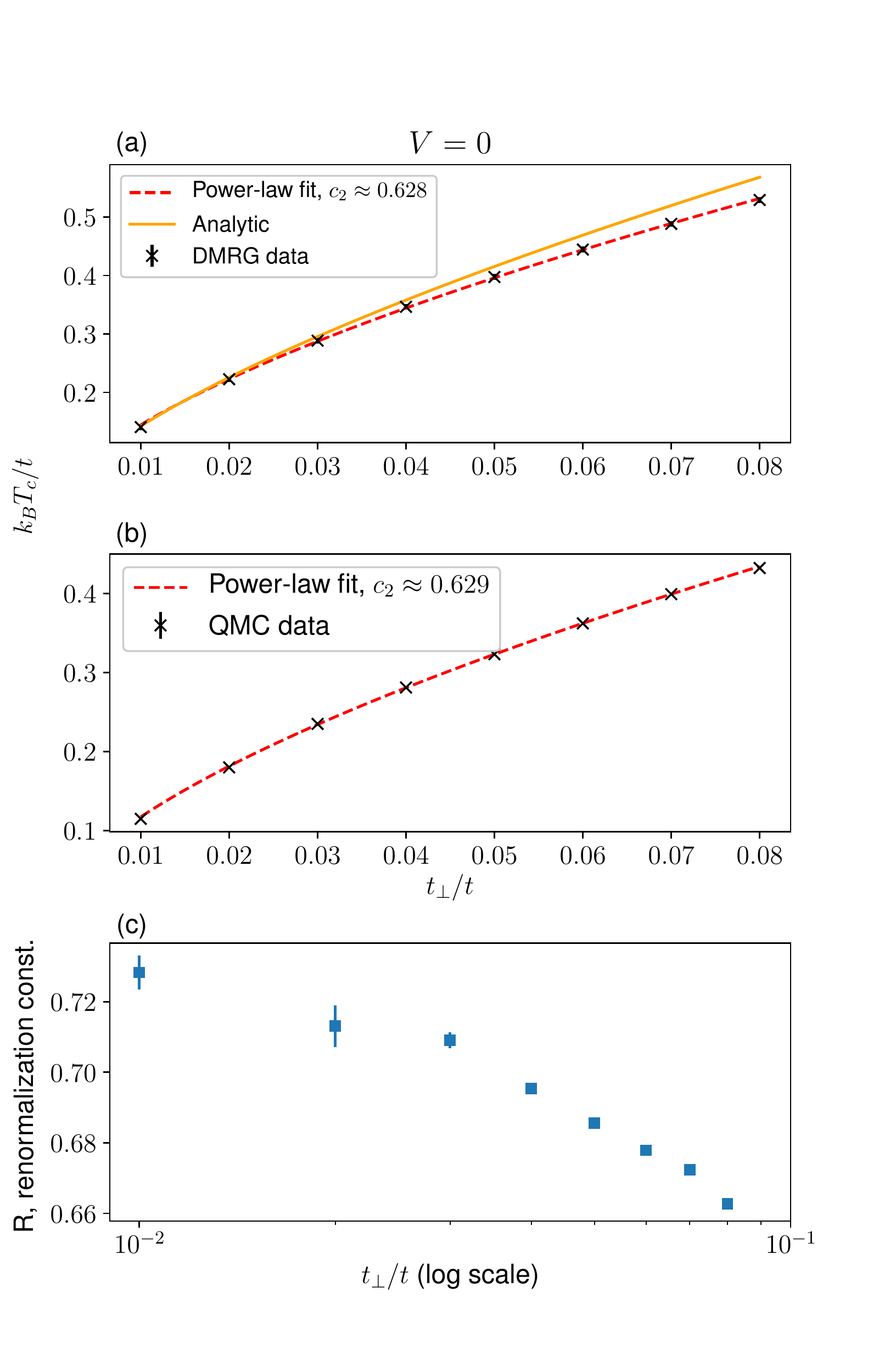}
	\caption{The critical temperature $T_c$ of hard-core bosons at $V=0$ and $n=0.5$ versus the transverse hopping $t_\perp$. (a) The red dashed line is a power-law fit to the data points (black crosses). The orange solid line is an analytical computation of $T_c$ based on Eq.~\eqref{Tc_equation}. (b) A power-law fit to the QMC $T_c$ data. (c) The constant $R$ from Eq.~\eqref{Tc_with_R} so that the QMC $T_c$ fits the analytical expression.}
	\label{Tvtperp_hcb}
\end{figure}

The dependency of $\langle b\rangle$ on $T$ does not change qualitatively with $t_\perp$, but the value of $T_c$ does scale with $t_\perp$, as shown in Fig.~\ref{Tvtperp_hcb}. 

Combining bosonization and mean field theory this scaling has been obtained as $T_c \sim t_\perp^{\frac{2}{3}}$ for this system~\cite{Cazalilla2006a}. Thus, we have performed a fit to the data with a power law given by 
\begin{equation} \label{T_c_scaling}
T_c = c_1t_\perp^{c_2}.
\end{equation}

In Fig.~\ref{Tvtperp_hcb}(a) we perform a power-law fit of our data and obtain the exponent $c_2\approx0.628$. The scaling disagrees somewhat with the analytical value of $c_2=2/3$. This is expected as the analytical value is less accurate for larger $T_c$. Further, we find the QMC $T_c$ scaling, $c_2\approx0.629$, by fitting to all data points in the same manner. We do not expect the scaling to agree with the analytical expression which relies on mean-field theory - see Sec.~\ref{sec:discussion} for a discussion of the different scaling behaviors.

Using the same approach it is also possible to produce an analytical expression for the critical temperature~\cite{Cazalilla2006a}:
\begin{equation} \label{Tc_equation}
	T_c = \frac{v_sn}{4\pi}\left[F(K)\frac{t_\perp z_c}{v_sn}\right]^{\frac{2K}{4K-1}},
\end{equation}
where $K,~v_s$ are the Tomonoga-Luttinger liquid (TLL) parameters~\cite{BookGiamarchi2003},  $n$ is the density and $z_c$ the coordination number. The function $F$ is given by
\begin{equation}
	F(K) = A_B(K)\sin\left(\frac{\pi}{4K}\right)\beta^2\left(\frac{1}{8K}, 1-\frac{1}{4K}\right),
\end{equation}
where the amplitude $A_B(K)$, relating the microscopic lattice operators to the ones of the effective field theory, is nonuniversal and depends on the specifics of the model, and $\beta(x,y)$ is the Euler beta function. Within the mean-field approximation, Eq.~\eqref{Tc_equation} is exact and fit free, as long as $K$, $v_s$, and $A_B$ are known. Hence, using ground-state DMRG we can obtain these three parameters from numerical fitting of the single particle density matrix~\cite{BookGiamarchi2003} at $T=0$. Thus, it is possible to produce critical temperatures given a ground-state calculation of a 1D system with conserved quantum numbers which is considerably less costly computationally. These values will be good approximations as long at $T_c$ is only a small fraction of the systems bandwidth - the deviations between Eq.~\eqref{Tc_equation} and our MPS+MF numerics at larger $T_c$ values visible in Fig.~\ref{Tvtperp_hcb}(a) are due to this. Conversely, at small $T_c$ the agreement is excellent.

It is possible to extract critical temperature dependence on $t_\perp$ from QMC as well and the results are shown in Fig.~\ref{Tvtperp_hcb}(b). Using the analytical expression~Eq.~\eqref{Tc_equation} with a renormalization of $t_\perp$ allows the overlapping of QMC data and analytical data~\cite{Yasuda2005}:
\begin{equation} \label{Tc_with_R}
	T_c = \frac{v_sn}{4\pi}\left[F(K)\frac{R\cdot t_\perp z_c}{v_sn}\right]^{\frac{2K}{4K-1}},
\end{equation}
where the renormalization constants $R\in[0.74,0.66]$, depending on $t_\perp$, is found to fit the QMC data as shown in Fig.~\ref{Tvtperp_hcb}(c). This renormalization constant has been discussed in the literature extensively~\cite{irkhin_calculation_2000,Yasuda2005,praz_scaling_2006,yao_universal_2007,thielemann_field-controlled_2009,furuya_dimensional_2016,blinder_nuclear_2017}, but here we find that as $t_\perp$ decreases, it appears to converge to a larger value than the one found in Ref.~\cite{Yasuda2005} for the case of an $SU(2)$-invariant system.

\subsubsection{$T_c$ dependence on $V$}
The data presented so far for finite temperature have been in the simplified regime of no nearest-neighbor repulsion $V=0$. However, the MPS+MF algorithm garners none or slight penalties in having finite $V$. This yields the possibility of measuring how the critical temperature depends on repulsive interactions. Further, it is interesting to see whether the relation between $T_c$ estimates from QMC and MPS+MF remains the same when interactions are turned on.

\begin{figure}[H]
	\includegraphics[width=\columnwidth]{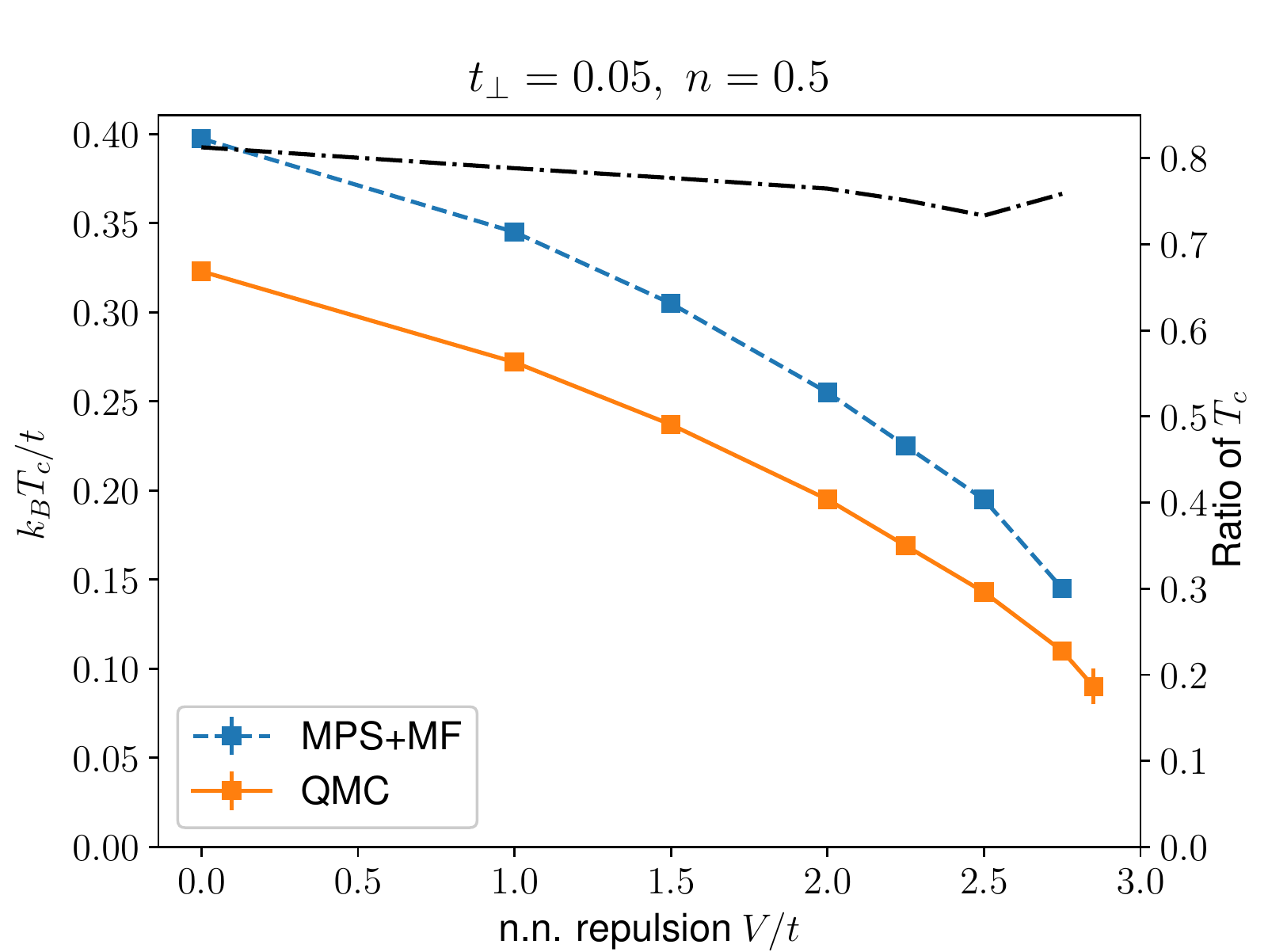}
	\caption{A plot of the critical temperature from superfluid to normal phase. The blue dashed line is computed using MPS+MF with OBC and the orange solid line using QMC. The black dash-dotted line is a ratio of the two results with values on the right axis.}
	\label{TcvV_tperp_005}
\end{figure}

In general, when repulsive interactions are turned on we know from Fig.~\ref{SF_CDW_trans} that the superfluid should weaken. We expect that the critical temperature is depressed for stronger interactions and this can be seen in Fig.~\ref{TcvV_tperp_005}. Remarkably, the ratio of critical temperatures is confined to a narrow band

\begin{equation}
	0.73 < \frac{T_c^{QMC}}{T_c^{MPS+MF}} < 0.82.
\end{equation}
This remains true even when the quantum critical point at $V_c/t\approx 3$ is approached.

We point out that the MPS+MF approach we have developed here has a crucial advantage over the TLL+MF framework behind Eq.~(\ref{Tc_equation}): It can compute $T_c$ even in regimes where the individual 1D systems no longer realize a TLL, such as for $V/t>2$ for HCBs and $U/t>3.3$ for SCBs, as shown in Figs.~\ref{orpvT_hcb}(b) and \ref{TcvV_tperp_005}.
\section{Discussion}\label{sec:discussion}
The zero-temperature SF-CDW transition is an example of so-called dimensional cross-over~\cite{Ho2004}. We can see this by noting that the order parameter for SF witnesses an exchange of bosons between chains. When the order parameter for SF goes to zero tunneling between chains is completely suppressed. The system now behaves more like a set of 1D systems with a remnant of interchain coupling only QMC can still resolve, whereas in the case of finite SF order parameter the exchange of particles made the system fully 3D. So, the cross-over from 3D to 1D becomes especially pronounced around the transition point. A major difference to the quantum phase transitions from SF to CDW occurring in 3D systems with isotropic tunneling and interactions that we are showing here is that the transition in the present quasi-1D systems is not second order, but first order for HCBs (see Appendix~\ref{first_order_motiv}), and possibly also for SCBs, as discussed below.

At the same time, it is evident from Fig.~\ref{SF_CDW_trans} that soft-core and hard-core bosons have qualitatively similar behavior. One notable difference is that the charge gap is much larger in the soft-core case. This is likely due to the transitions occurring at much larger values of $U$. Thus, the energy penalty for adding and removing particles is much larger than in the hard-core case. In addition, there is a small region around the transition where the two order parameters coexist, i.e., where both are small but finite. But as the limitations of mean-field approaches (of which we are using a partial one) in predicting supersolids are well documented, we refrain from concluding the existence of such a state here. We also note that the SF order parameter is enhanced for the soft-core case. This could be explained by the fact that sites are almost never locked as they often would be in the hard-core case. In other words, it is almost always possible to inject particles into the system in contrast to the hard-core case. If the maximum boson number is reached the site is artificially locked but the amplitude for a state where this occurs is negligible in accordance with~Eq.~\eqref{projection}. 

For soft-core bosons, in the true 1D case the CDW transition occurs at $U_{1\text{D}}/t\approx 3.3$~\cite{krutitsky_ultracold_2016} whereas in the quasi-1D case with $t_\perp/t=0.05$ (see Fig.~\ref{SF_CDW_trans}) it does not occur until $U_{\text{Q}1\text{D}}/t\approx8.12$~\cite{Kuhner2000a}. For comparison, a 3D system with isotropic tunneling yields $U_{3\text{D}}/t=29.94(2)$~\cite{Capogrosso-Sansone2007} indicating that the large increase we observe from $U_{1D}$ to $U_{Q1D}$ is reliable and the value of $U_{Q1D}$ is heavily dependent on $t_\perp$.

Overall, the soft-core boson case appears to differ qualitatively from the hard-core case when it comes to transition order. In the soft-core model we could not find any clear first order behavior in the superfluid order parameter while the charge gap behaves similarly to the hard-core case - in fact, charge gaps in both cases show a more pronounced discontinuity than the supefluid order parameter or the transverse superfluid stiffness~(see Fig.~\ref{SF_CDW_trans}). While we cannot detect a jump in the SF order parameter, and thus a full first-order transition, for the SCB system, this scenario remains the most likely explanation for the observed behavior. We note that there may be effects of the mean-field approximation that degrade any jump below the threshold that we could numerically resolve. The only other alternative we see that could explain the behavior of Fig.~\ref{SF_CDW_trans} is that of a simultaneous SF and CDW order, i.e., a supersolid. As discussed above, it appears to us that such an alternative would however require more evidence than what can be supplied with the MPS+MF approach on its own.

For the critical temperatures of the SF to normal transition, the analytical prediction agrees well with the numerical correspondents as seen in Fig.~\ref{Tvtperp_hcb}, especially at low $T_c$'s. It is notable that the scaling of both MPS+MF and QMC data are very similar, another positive for the approximative MPS+MF approach, with a power below that of the power of $2/3$ predicted from TLL+MF. At small $t_\perp$ we expect and find improved agreement between analytical theory and MPS+MF, in line with the fact that the TLL+MF prediction will work better as $T_c$ becomes a small fraction of the systems bandwidth. For both the MPS+MF and the QMC data we find that increasingly constraining the fitting window to the smallest values of $t_\perp$ yields exponents approaching $c_2\approx2/3$ from below, showing that the mean-field approximation becomes better with decreasing $t_\perp$. The close agreement in the scaling behavior of $T_c$ with $t_\perp$ between the QMC and MPS+MF techniques, and their common disagreement with the $t_\perp^{2/3}$-scaling derived from TLL+MF points to the source being within the TLL approximation of the microscopic lattice Hamiltonian of the chains.

For the temperature data at finite repulsion in a hard-core system it is interesting to note the relative constancy of the $T_c$ ratio between MPS+MF and QMC. The different critical temperatures seem to agree less for larger values of repulsion with the exception of the point at $V=2.75$ where there is a different trend.

While the results between QMC and MPS+MF differ somewhat we note the differing efficiency of the two algorithms. QMC data in this paper have been obtained using 30000 equilibration steps and 1000000 measurement steps. For a single core [Intel(R) Xeon(R) Gold 6140 CPU @ 2.30GHz] we find that
\begin{itemize}
	\item CPU time for equilibration $\approx 0.001L^4$ sec
	\item CPU time for measurements $\approx 0.1L^4$ sec
\end{itemize}
For MPS+MF we note that scaling is exactly that of typical DMRG:
\begin{equation}\label{dmrg_scaling}
	t_{tot} \sim N_{sol}d^2\chi^3L,
\end{equation}
where $d$ is the local Hilbert space dimension, $\chi$ is the bond dimension, and $L$ the system size. The MPS+MF routine has the added complication of having to perform several DMRG calculations. We have found that the number of required solutions, $N_{sol}$ vary greatly, particularly close to transitions. Deep in an ordered phase the number of required solutions can be as low as $N_{sol}\sim5$. Close to a phase transition we find this number able to reach $N_{sol}\sim50$ for OBC and $N_{sol}\sim30$ for PBC including the various guessing heuristics we employ as mentioned in Sec.~\ref{sec:methods}. Most important is that $N_{sol}$ is not very dependent on system size, approximately conserving the L dependence of Eq.~\eqref{dmrg_scaling}.

We compare data using a PBC model and the efficiency should be compared between these two cases as well. Note that using OBC gives an incredible boost to efficiency due to the lower bond dimension, which can be used for the finite temperature case. It is further worthwhile to note that QMC would obtain a better scaling with system size for finite temperature and thus shorter run times as well.

On an [Intel(R) Xeon(R) Processor E5-2630 v4 CPU @ 2.20GHz] we find an $L=60$ system running for $\approx1600$ seconds per solution in MPS+MF. With the largest number of loops at $N_{sol}\sim30$ we arrive at
\begin{itemize}
	\item QMC time: 15 days
	\item MPS+MF: 0.55 days.
\end{itemize}
As expected the MPS+MF algorithm ends up comparing well when doing single-core calculations. It is worthwhile to mention that QMC can scale up its measurement phase to several cores where calculation speed increases linearly with each core added. The degree with which MPS-based codes can exploit parallelism  varies widely by implementation, but linear speedups in the number of CPU cores are generally not available over as wide a range as for QMC. Nevertheless, scientific projects typically address some finite area of parameter space, meaning that the MPS+MF can obviously exploit perfect (and trivial) parallelism in system parameters.

\section{Conclusion}\label{sec:conclusion}
Our results show that an approach using DMRG to solve a decoupled 3D system self-consistently is valid for use on an anisotropic system and also reproduces the transition points with reasonable accuracy. In particular, the SF to CDW phase transition has nearly equal critical repulsion $V_c$ for the MPS+MF case compared to QMC. The major benefit is that the DMRG approach is computationally cheaper than the corresponding exact QMC. We will further be able to simulate real-time dynamics on the states produced by this framework of MPS+MF. For the finite temperature transition to a normal phase the critical temperature deviates more from the exact case. However, this deviation is much less sizable than what a full mean-field approximation produces. This method presents a powerful possibility of treating anisotropic 2D and 3D systems quickly using DMRG, in particular beyond the TLL approach.

Another key finding of this work is the first-order nature of the quantum phase transition between the superfluid and the charge density wave order for hard-core bosons in these quasi-1D anisotropic systems, as opposed to the expected purely second order transition in a 3D system isotropic in tunneling (and interactions, in the case of HCBs). At the same time, the discontinuous opening of the charge gap contra the apparent continuous vanishing of the SF order parameter, which occurs for the case of soft-core bosons, may indicate different orders of the transition in that specific system. The former suggesting first order while the latter looks like second order. Our current method and analysis is insufficient to determine whether there is a very small jump. If that is the case it is further possible that the gap gets smoothed out by the mean-field treatment. A more detailed analysis of the soft-core model is required to ascertain whether the transition is truly first order.

The method presented in this paper reproduces previous analytical results. Critical temperature calculations using this method scale with transverse hopping strength $t_\perp$ corresponding to what you would obtain using an effective field theory on the 3D system and then decoupling with mean-field theory. Replacing the effective field theory with DMRG we find similar scaling laws with a modified exponent. In addition, using ground state data from the normal 1D MPS routine we may produce a critical temperature estimate from the field theory. Both the scaling and estimated value agree well with the presented approach at small $t_\perp$, where agreement is expected. The reasonably close agreement to theory allows us to trust our numerical methods in the context of mean-field theory. Combining analytical and numerical methods in this manner could allow us to obtain $T_c$ estimates in parameter regions that are too computationally costly. This will be especially true for an extension of our method to fermionic systems, where, even putting aside the sign problem, auxiliary-field QMC approaches scale much worse in the number of lattice sites than in QMC for bosons (cubic vs linear scaling).

\begin{acknowledgments}
	We would like to thank Thierry Giamarchi, Thomas K\"ohler and Shintaro Takayoshi for useful discussions. This work has received funding through an ERC Starting Grant from the European Union's Horizon 2020 research and innovation programme under Grant agreement No. 758935. Part of the simulations were performed using the MPToolkit, written by Ian McCulloch \cite{MPTK_webpage}. The computations were performed on resources provided by SNIC through Uppsala Multidisciplinary Center for Advanced Computational Science (UPPMAX) under Projects SNIC 2019/3-323, SNIC 2019/8-26,	SNIC 2020/1-48, SNIC 2019/35-16, uppstore2019070, and	SNIC 2020/16-92.
	
	NL thanks the French National Research Agency (ANR) for support under project THERMOLOC ANR-16-CE30-0023-0. We acknowledge CALMIP (Grant Nos. 2018-P0677 and 2019-P0677) and GENCI (Grant A0030500225) for high-performance computing resources.
\end{acknowledgments}

\appendix

\section{OBC boundary contamination} \label{App_OBC}
When using OBC the boundaries are dissimilar from other sites in the system in that they are missing one neighboring site. Depending on the Hamiltonian this causes a bias towards either holes or particles to occupy the edge sites.

\begin{figure}[H]
	\includegraphics[width=\columnwidth]{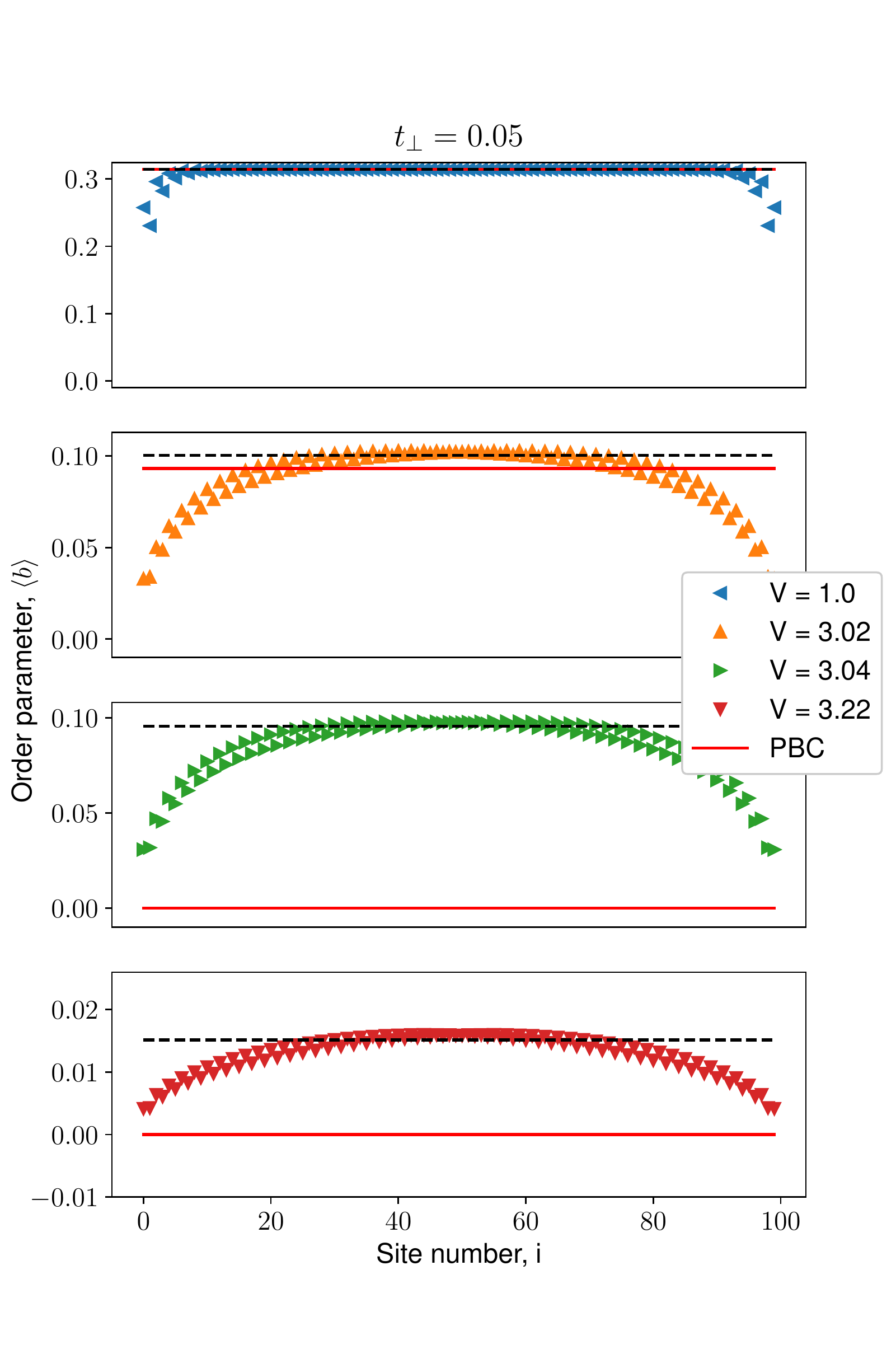}
	\caption{Order parameter $\left<b\right>$ across the hard-core system for different values of nearest neighbor repulsion $V$ using OBC. The OBC data is for an $L=100$ system and the PBC data is extrapolated to $L\to\infty$.}
	\label{HCB_OBC_orderpar}
\end{figure}

The usual method to deal with this bias is to focus on the central part of the system and assume that boundary effects do not reach in beyond a certain point. However, as can be seen from Fig.~\ref{HCB_OBC_orderpar}, the assumption does not hold for the case of the Hamiltonian in Eq.~\eqref{smf_ham}.
Instead we see the boundaries start a pattern of alternating particles and holes. Since it is clearly preferable to have particles on the edges in the considered system, the two edges have large weight on the occupation state. After $V/t=3$ the pattern becomes increasingly apparent and finally the average simply does not attain the system center value. Further, even if the average was a good measure of the center value it can be seen from Fig.~\ref{HCB_OBC_orderpar} that the boundaries actually incur a finite superfluid order inside the system which leads to a transition occurring only at $V_c/t=3.24$.

It is further clear that this is a boundary effect since when system size is increased above the sizes used in this paper the finite size trend changes, making extrapolations difficult. In practice, to overcome the boundary effect on superfluid order in the hard-core system with nearest neighbor interaction we find that sizes of $L=200$ are insufficiently long to see any convergence.

This clearly shows the periodic boundary conditions are necessary to analyze the hard-core bosons with nearest neighbor repulsion since extrapolations to infinite size suffer no trend changes at moderate sizes.

\begin{figure}[H]
	\includegraphics[width=\columnwidth]{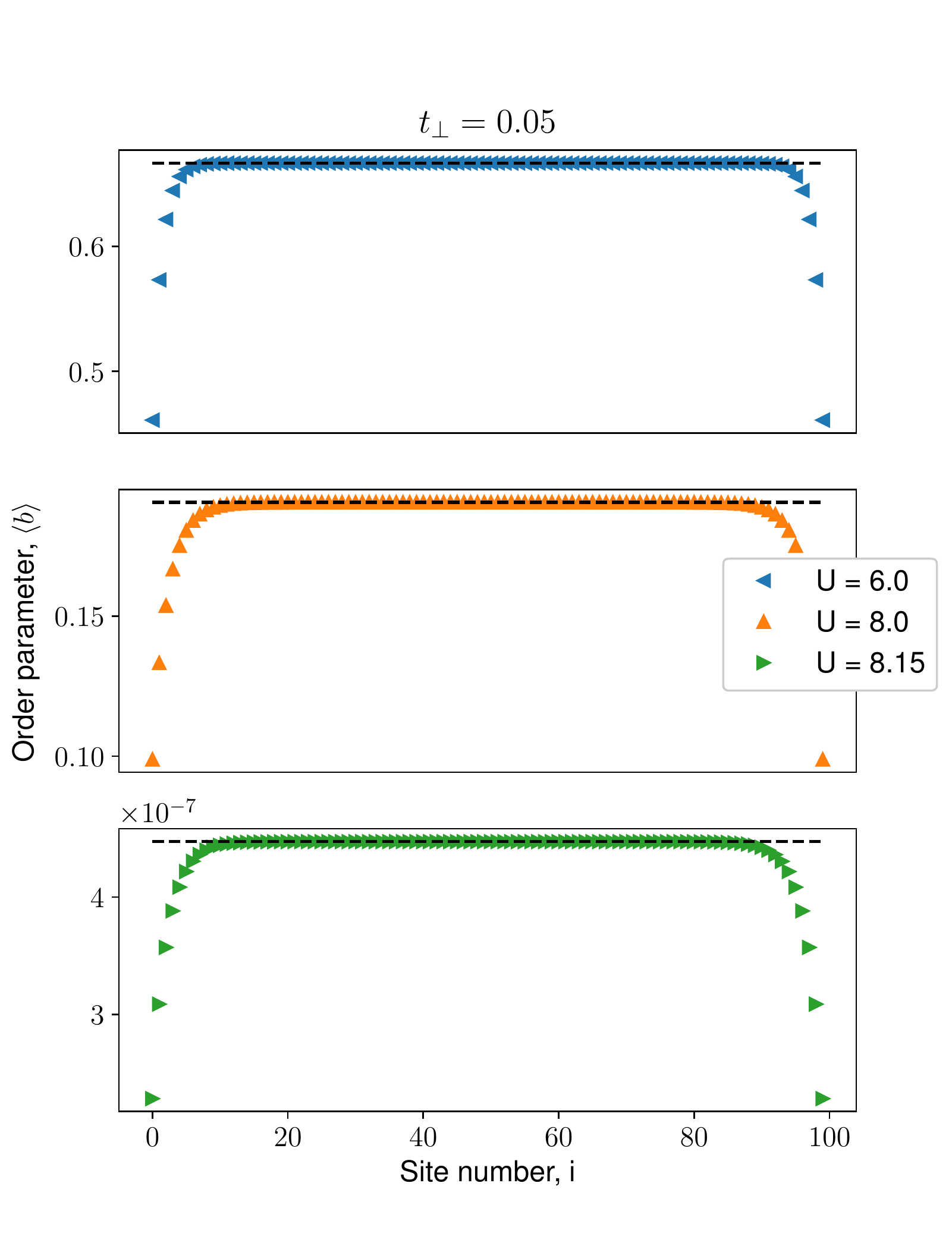}
	\caption{Order parameter $\left<b\right>$ across the soft-core system for different values of onsite repulsion $U$ using OBC.}
	\label{SCB_OBC_orderpar}
\end{figure}

This is a much smaller problem in the case of soft-core bosons with onsite repulsion as seen in Fig.~\ref{SCB_OBC_orderpar}. We can clearly see the order parameter saturate to a specific value at the center of the system quite quickly. Further, the plateauing does not seem to be strongly affected by the onsite repulsion.
Evidently, when there is no nearest neighbor interaction the effect of the boundaries is much smaller and OBC can safely be used. Due to these observations we assume that PBC will not yield a different result than OBC and neglect to perform the costly computations soft-core bosons with PBC would entail.

Thus we have chosen to use PBC for the hard-core system with nearest neighbor interactions and OBC for the soft-core system with onsite interactions.

\section{Density plateaus} \label{App_plateau}
\begin{figure}[H]
	\includegraphics[width=\columnwidth]{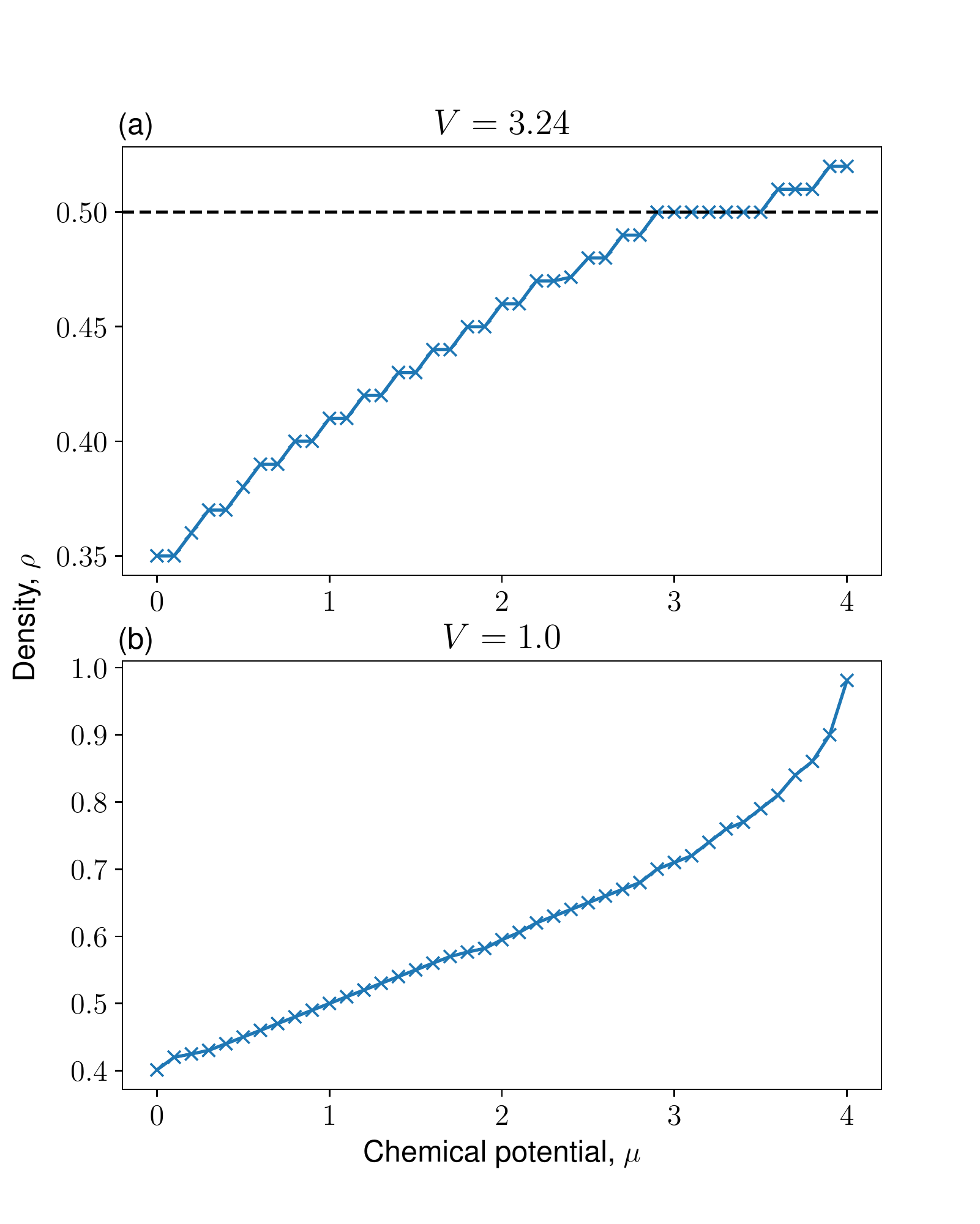}
	\caption{Showing the density of a system with $\alpha=0$ at a $V$ in the (a) CDW phase and (b) the TLL phase. Note the clear plateau around $n=0.5$ inside the CDW phase.}
	\label{plateau}
\end{figure}

When the considered model does not conserve particle number it is not possible to use energy differences of states with different particle number to determine the charge gap (as in, e.g., Karakonstantakis \textit{et al.}~\cite{Karakonstantakis2011}). This is because it can occur that
\begin{equation} \label{densofmu}
n\left(\mu\right) = n\left(\mu+\delta \mu\right),
\end{equation}
where $\delta\mu$ is some small shift from $\mu$. In practice, this occurs in the CDW phase which yields a certain arbitrariness to the energy since certainly
\begin{equation}
E(\mu) \not= E(\mu+\delta\mu)
\end{equation}
as long as there are any particles in the system, while from Eq.~\eqref{densofmu} we would obtain
\begin{equation}
E(N)=E\left(n\left(\mu\right)\right) =  E\left(n\left(\mu+\delta \mu\right)\right).
\end{equation}
Another method may be used based on the variation of $\mu$. When computing density versus chemical potential, in the CDW phase you find plateaus of constant density, as shown in Fig.~\ref{plateau}(a), whose widths are the energies required to increase particle number by one. It is possible to compute how much the chemical potential $\mu$ must be increased (decreased) to obtain an increase (decrease) in the systems density. This yields an upper and lower chemical potential for that particular density. The difference of this upper and lower bound is then the energy required to increase/decrease particle number.

The width of the density plateau $W$ is related to an energy difference obtained from a number-conserving calculation once you enter the CDW phase of the system
\begin{equation}
W \approx E(N+1) + E(N-1) - 2E(N).
\end{equation}
Further, as can be seen from Fig.~\ref{plateau}(b), when repulsion is decreased and we enter the superfluid phase the plateauing tendency disappears.
For charge gaps computed in the paper we have used a precision which is at worst $\epsilon_\mu=1e^{-3}$ for the upper and lower limit of the plateau.

\section{Isotropic tunneling} \label{iso_example}
The first-order nature of the transition is not entirely clear in the QMC data in Fig.~\ref{SF_CDW_trans}. This is due to the fact that $t_\perp=0.05t$ is very small. To elucidate the nature of the transition we may consider larger values of $t_\perp$ as we still expect the system to be in the same universality class.

\begin{figure}[H]
	\includegraphics[width=\columnwidth]{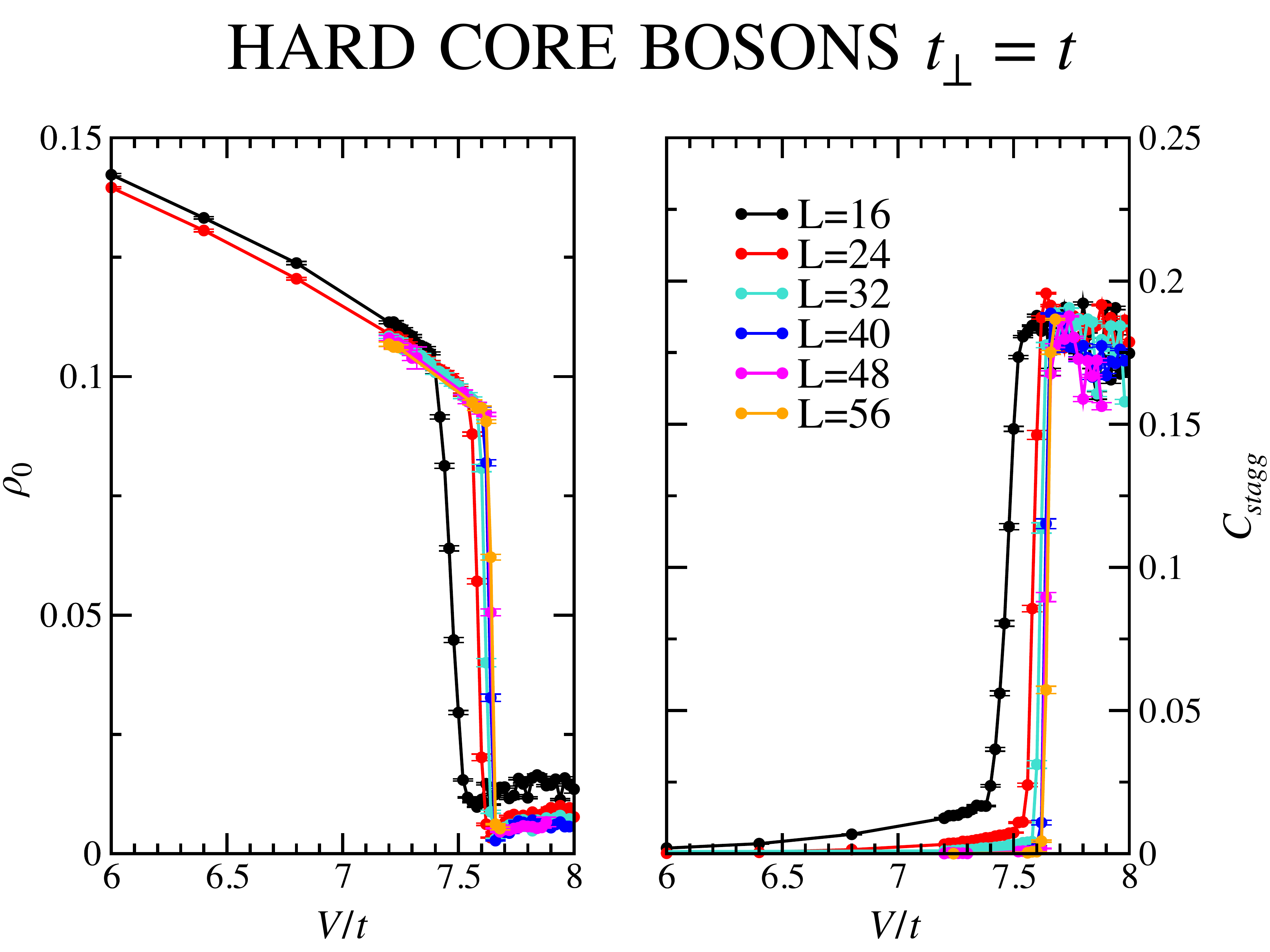}
	\caption{BEC order parameter and CDW order parameter for the isotropic case $t_\perp=t$.}
	\label{SF_ISO}
\end{figure}

For an isotropic case of $t_\perp=t$ the correspondent result of Figs.~\ref{SF_CDW_trans}(b) and \ref{SF_CDW_trans}(c) is given in Fig.~\ref{SF_ISO}. Both order parameters tend to exhibit clear jumps that become sharper as system size is increased. We thus reason that the gap should remain, albeit diminished, in the case of anisotropic tunneling.

\section{First order transition} \label{first_order_motiv}
The jump in the order parameters given by the MPS+MF routine in Fig.~\ref{SF_CDW_trans}(a) and Fig.~\ref{SF_CDW_trans}(c) do not by themselves guarantee first-order behavior. For the latter one of the order parameters seems to vanish continuously as far as we can resolve.

To clarify the transition order we compute additional indicators. For the soft-core case we find no issue in computing correlation functions and obtaining correlation lengths using the scaling behavior for single-particle density matrix and density-density correlator respectively:
\begin{equation} \label{1pdm_scale}
	\braket{b^\dagger_ib_{i+r}} \sim A_0 + B_0\exp(-r/\xi),
\end{equation}
\begin{equation} \label{dd_scale}
	\braket{n_in_{i+r}} \sim A_1 + B_1\exp(-r/\xi),
\end{equation}
where $\xi$ is the correlation length (differing between the two correlators).

\begin{figure}[H]
	\includegraphics[width=\columnwidth]{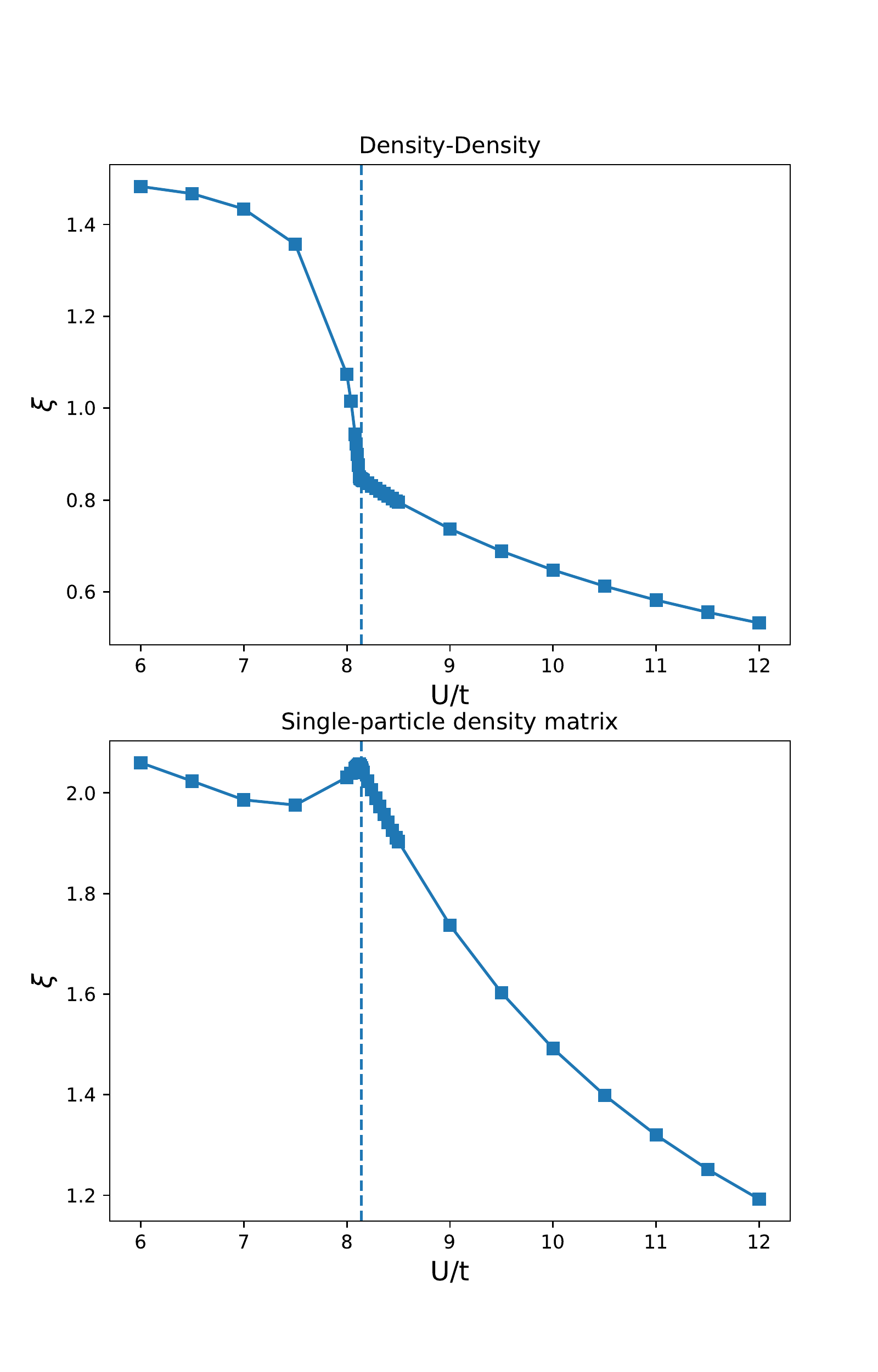}
	\caption{Density-density correlator and single-particle density matrix versus onsite repulsion $U/t$ in an $L=100$ system for transverse hopping $t_\perp=0.05t$ at bond dimension $\chi=50$.}
	\label{SCB_corr_len}
\end{figure}

As can be seen from Fig.~\ref{SCB_corr_len} there is a change of trend in the correlation length at the transition. In addition, the single-particle density matrix has an increasing correlation length with a maximum at the transition. As far as we can resolve there is no divergence and no increased tendency thereof with increased bond dimension. This indicates we are capturing the correct behavior. Since the correlation length is finite all across the transition we conclude that the soft-core superfluid to charge density wave transition is first order.

For the hard-core system it is difficult to fit Eq.~\eqref{1pdm_scale} and \eqref{dd_scale} to the measured correlators. We find a nonexponential trend taking over after a short distance suggesting insufficient bond dimension for carrying the correlations over sufficient distance to obtain good fits for a correlation length.

To obtain additional proof of the transition order in this case we instead measure the (infinite size) gap to the first excited state
\begin{equation} \label{sol_gap}
	\Delta_s = \lim\limits_{L\to\infty}E_1(L) - E_0(L).
\end{equation}
For a second order transition we would expect the gap to the first excited state to be unchanged across the transition. 
\begin{figure}[H]
	\includegraphics[width=\columnwidth]{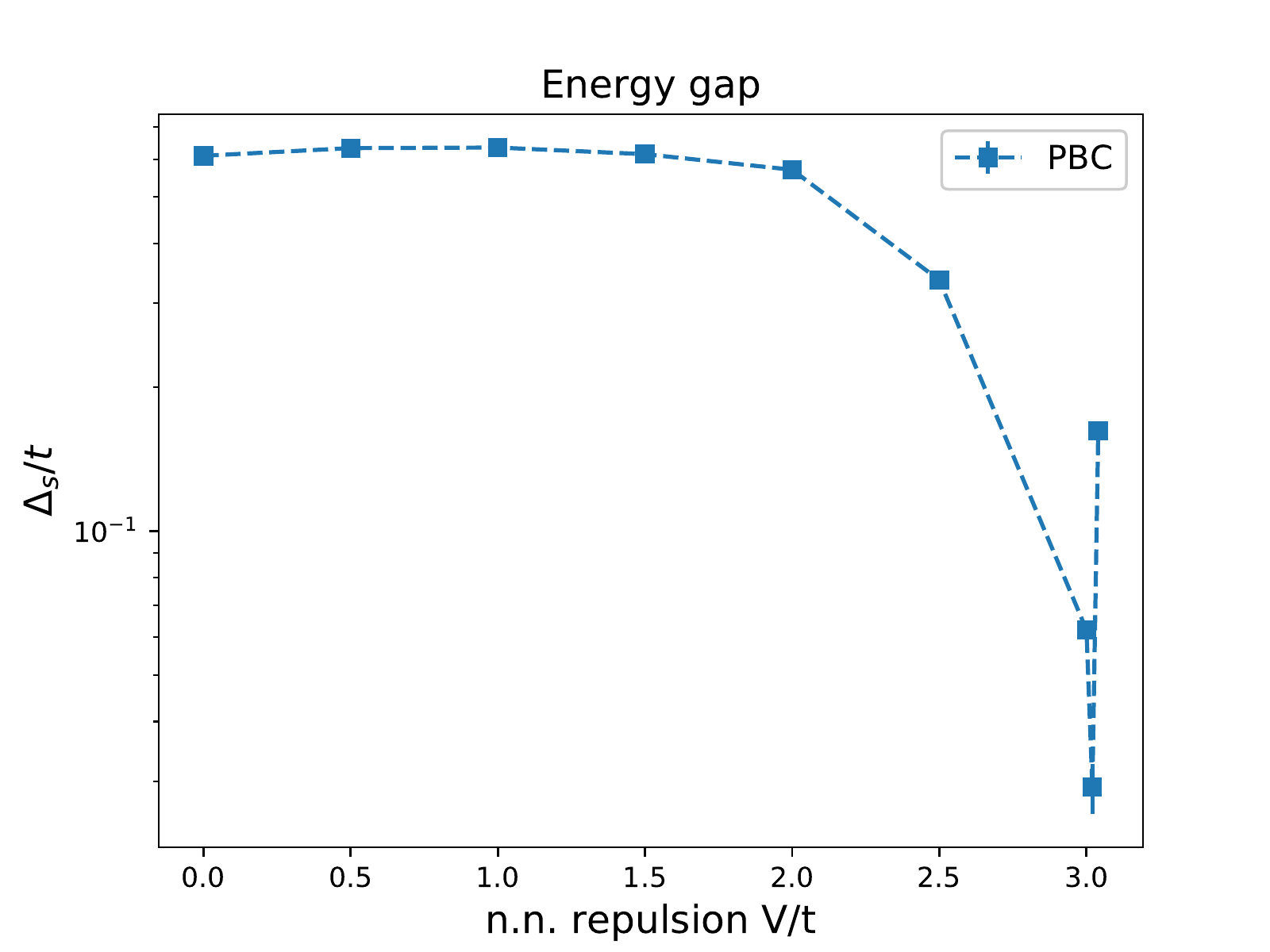}
	\caption{Energy gaps to the first exited state extrapolated to infinity using PBC for bond dimension $\chi=250$ with transverse hopping $t_\perp=0.05t$.}
	\label{sol_gaps}
\end{figure}

As can be seen from Fig.~\ref{sol_gaps} the gap defined by Eq.~\eqref{sol_gap} jumps by an order of magnitude across the transition and changes direction abruptly. We take this to indicate a first order transition since the gap to the excited state is not changing smoothly over the transition. Together with the jumplike behavior of both the SF and CDW order parameter in Fig.~\ref{SF_CDW_trans}(a) we conclude that the hard-core transition is first order as well.

\section{Finite-size extrapolation} \label{App:fss}
\begin{center}
	\begin{figure}[b]
		\includegraphics[width=\textwidth]{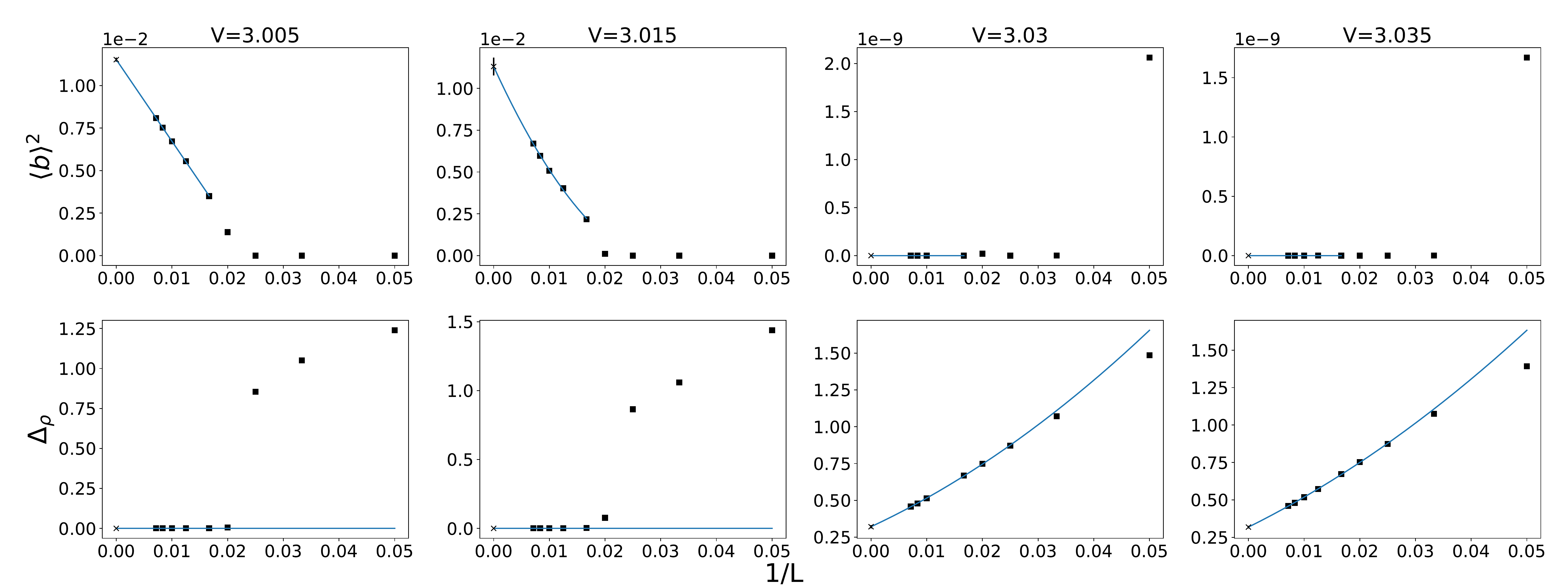}
		\caption{Squared order parameter and charge gap versus inverted system size (black squares) for the hard-core boson model with PBC. The blue solid line is a fit following Eq.~\eqref{orp_fit_2} for the order parameter and Eq.~\eqref{cgap_fit} for the charge gap.}
		\label{fss_hcb_pbc}
	\end{figure}
\end{center}

We present the finite size data which after extrapolation yields the data given by MPS+MF in Fig.~\ref{SF_CDW_trans}(a)~and~\ref{SF_CDW_trans}(d). The finite size data has been computed for a varying range of sizes depending on when a clear trend which could be extrapolated appeared. Results for the hard-core model are shown in Fig.~\ref{fss_hcb_pbc}. Notably, on the superfluid side order parameter increases with size. In addition, there is a certain size at which the system no longer supports superfluidity (e.g., $L=40$ for n.n. repulsion $V=3.005$). With increasing repulsion larger sizes are required to obtain superfluidity. We find that after $V=3.02$ no sizes manage to obtain superfluidity and all order parameters are zero.

For the charge gap similar conditions hold. We find that below transition smaller sizes obtain a finite charge gap whereas in larger systems it is consistently zerovalued. This comes as no surprise since when superfluidity fails the system is truly 1D and should follow such physics. In this case, a truly 1D system at these parameters transitions to a CDW phase at $V=2$ and we find the expected charge gaps at sizes where superfluidity disappears.

\onecolumngrid

\begin{center}
	\begin{figure}
		\includegraphics[width=\textwidth]{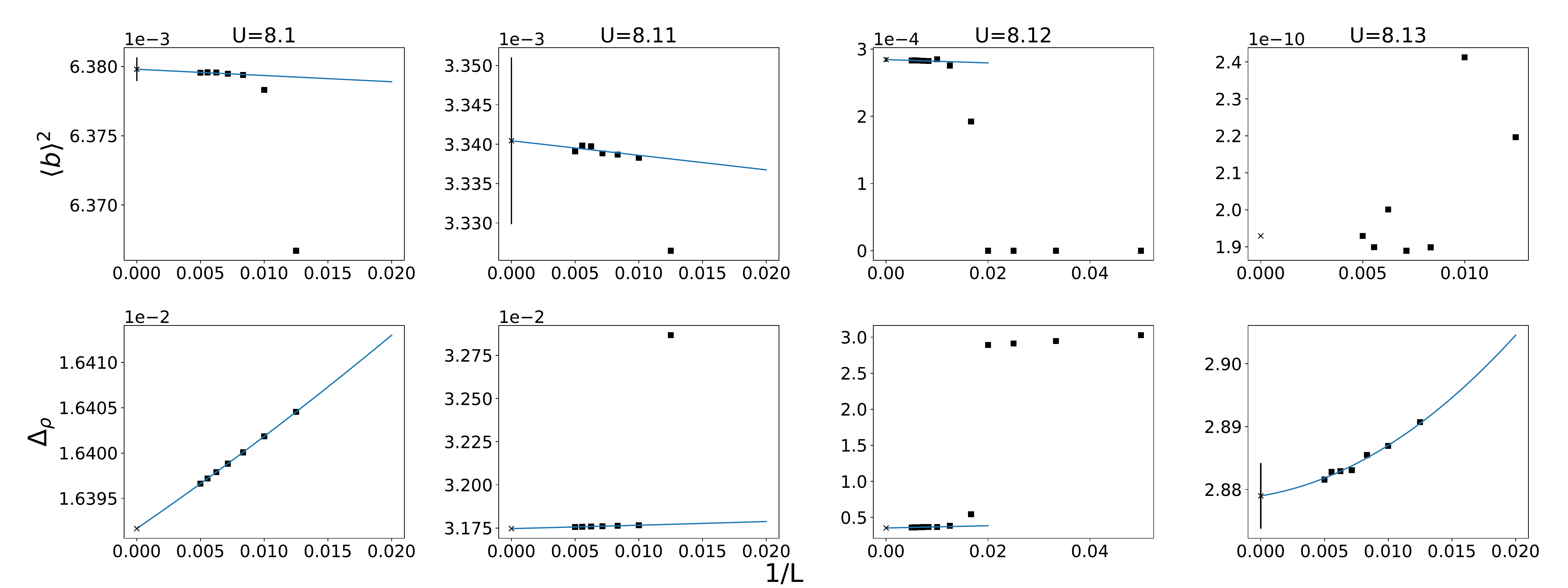}
		\caption{Squared order parameter and charge gap versus inverted system size (black squares) for the soft-core boson model with OBC. The blue solid line is a fit following Eq.~\eqref{orp_fit_2} for the order parameter and Eq.~\eqref{cgap_fit} for the charge gap.}
		\label{fss_scb}
	\end{figure}
\end{center}
\twocolumngrid

For the soft-core case results look somewhat different as shown in Fig.~\ref{fss_scb}. This is mainly due to OBC allowing us access to much larger system sizes such that only systems with finite order parameter have been considered on the superfluid side. We note that the order parameter seems to obtain large fitting errors. In relation to the size of the order parameter these errors typically remain on the order of marker size and are included in Fig.~\ref{SF_CDW_trans}(d).

For the charge gap we note that it looks finite before transition. This is within the error produced by our charge gap routine as outlined in Appendix~\ref{App_plateau} and we consider these charge gaps zerovalued. At onsite repulsion $U=8.12$ we find the curious case of simultaneous finite (but small) charge gap and order parameter despite extrapolation which is discussed in Sec.~\ref{sec:discussion}.

\bibliography{hcb_dmrg_mf,nicolas_refs}

\end{document}